\documentclass[12pt]{article}

\usepackage{arxiv}

\usepackage[utf8]{inputenc} 
\usepackage[T1]{fontenc}    
\usepackage{hyperref}       
\usepackage{url}            
\usepackage{booktabs}       
\usepackage{amsfonts}       
\usepackage{nicefrac}       
\usepackage{microtype}      
\usepackage{lipsum}
\usepackage{setspace}
\usepackage{tabulary}

\usepackage[pdftex]{graphicx}
\usepackage{subcaption}
\usepackage{epstopdf}
\usepackage{amssymb}
\usepackage[cmex10]{amsmath}
\usepackage{subfloat}
\usepackage{verbatim}
\usepackage{cite}
\usepackage{multirow}
\usepackage{multicol}
\usepackage{bm}
\usepackage{enumerate}
\usepackage{mathrsfs}
\usepackage{algorithm}
\usepackage{algpseudocode}
\usepackage{adjustbox}
\usepackage{mwe}
\usepackage{xr}
\usepackage{caption} 
\usepackage[title]{appendix}

\algdef{SE}[DOWHILE]{Do}{doWhile}{\algorithmicdo}[1]{\algorithmicwhile\ #1}%
\algnewcommand{\var}{\textit}

\graphicspath{{./figures/}{./plots/}}

\makeatletter
\let\OldStatex\Statex
\renewcommand{\Statex}[1][3]{%
	\setlength\@tempdima{\algorithmicindent}%
	\OldStatex\hskip\dimexpr#1\@tempdima\relax}
\makeatother

\newtheorem{Def}{\textbf{Definition}}

\hypersetup{
	colorlinks,
	citecolor=black,
	filecolor=black,
	linkcolor=black,
	urlcolor=black
}

\title{A Single SMC Sampler on MPI that Outperforms a Single MCMC Sampler}

\author{
	Alessandro~Varsi\\
	Department of Electrical Engineering and Electronics\\
	University of Liverpool\\
	Liverpool, L69 3GJ, UK \\
	\texttt{Alessando.Varsi@liverpool.ac.uk} \\
	\And
	Lykourgos~Kekempanos\\
	Department of Electrical Engineering and Electronics\\
	University of Liverpool\\
	Liverpool, L69 3GJ, UK \\
	\texttt{L.Kekempanos@liverpool.ac.uk} \\
	\And
	Jeyarajan Thiyagalingam \\
	Scientific Computing Department\\
	Rutherford Appleton Laboratory, STFC\\
	Didcot, Oxfordshire, OX11 0QX, UK \\
	\texttt{t.jeyan@stfc.ac.uk} \\
	\And
	Simon~Maskell\\
	Department of Electrical Engineering and Electronics\\
	University of Liverpool\\
	Liverpool, L69 3GJ, UK \\
	\texttt{S.Maskell@liverpool.ac.uk}\\
}

\begin{document}
\maketitle

\begin{abstract}
	Markov Chain Monte Carlo (MCMC) is a well-established family of algorithms which are primarily used in Bayesian statistics to sample from a target distribution when direct sampling is challenging. Single instances of MCMC methods are widely considered hard to parallelise in a problem-agnostic fashion and hence, unsuitable to meet both constraints of high accuracy and high throughput. Sequential Monte Carlo (SMC) Samplers can address the same problem, but are parallelisable: they share with Particle Filters the same key tasks and bottleneck. Although a rich literature already exists on MCMC methods, SMC  Samplers are relatively underexplored, such that no parallel implementation is currently available. In this paper, we first propose a parallel MPI version of the SMC Sampler, including an optimised implementation of the bottleneck, and then compare it with single-core Metropolis-Hastings. The goal is to show that SMC Samplers may be a promising alternative to MCMC methods with high potential for future improvements. We demonstrate that a basic SMC Sampler with $512$ cores is up to $85$ times faster or up to $8$ times more accurate than Metropolis-Hastings.
\end{abstract}


\section{Introduction} \label{sec:intro}
	In Bayesian statistics, it is often necessary to collect and compute random samples from a probability distribution. Markov Chain Monte Carlo (MCMC) methods are commonly used to address this problem since direct sampling is often hard or impossible. Sequential Monte Carlo (SMC) Samplers are a member of the broader class of SMC methods (which also includes Particle Filters) and can be used in the same application domains as MCMC\cite{delmoral}. While many papers on Particle Filters or MCMC methods exist, SMC Samplers still remain relatively unexplored as a replacement to MCMC.
	
	Research has been focused on improving the run-time and accuracy of SMC and MCMC methods to meet the constraints of modern applications. While accuracy has been improved by several approaches ranging from using better proposal distributions \cite{MH_better_proposal} to better resampling and better recycling \cite{septier}, to improve the run-time these algorithms need to employ parallel computing.
	
	Generic MCMC methods are not parallelisable by nature as it is hard for a single Markov chain to be processed simultaneously by multiple processing elements. In \cite{parallel_mcmc_prefetching1}, an approach which aims to parallelise a single chain is presented but it quickly becomes problem-specific because the efficiency of parallelisation is not guaranteed, especially for computationally cheap proposal distributions. We acknowledge that one could implement multiple instances of MCMC in parallel as in \cite{parallel_mcmc_2009}, but argue that we could also apply the same idea to multiple instances of SMC samplers. However, all chains also need to burn-in concurrently, making it difficult to use this approach to reduce the run-time. In this paper, we seek to develop a parallel implementation of a single instance of a sampling algorithm which outperforms a single MCMC algorithm both in terms of run-time and accuracy. Therefore, we leave comparisons with multiple-chain MCMC to future work along with comparisons with parallel instances of SMC Samplers.
	
	Particle Filters offer inherent parallelism, although an efficient parallelisation is not trivially achievable. The resampling step, which is necessary to respond to particle degeneracy \cite{Simon_1st}, is indeed a challenging task to parallelise. This is due to the problems encountered in parallelising the constituent redistribute step. Initial approaches to performing resampling are explained in \cite{Simon_1st}\cite{gordon} and achieve $O(N)$ time complexity. In \cite{Simon_2nd}, it has been proven that redistribute can be parallelised by using a divide-and-conquer approach with time complexity equal to $O((log_2N)^3)$. This algorithm has been optimised and implemented on MapReduce in \cite{Lykourgos} and then ported to Message Passing Interface (MPI) in \cite{Alessandro}. Although the time complexity is improved to $O((log_2N)^2)$, it has been shown that at least $64$ parallel cores are required to outperform the $O(N)$ redistribute version when all other steps are parallelised using MPI.
	
	No parallel implementation of the SMC Sampler on MPI is currently available, despite its similarities with the Particle Filter. Hence, the first goal of this paper is to show that an MPI implementation of the SMC Sampler can be translated from the MPI Particle Filter in \cite{Alessandro} by porting its key components. An optimisation of the redistribute in \cite{Alessandro} will also be discussed and included in the proposed algorithm. This paper also compares, both in terms of run-time and accuracy, a basic implementation of the SMC Sampler on MPI with an equally simple MCMC method, Metropolis-Hastings \cite{Metropolis1953}. By proving that the SMC Sampler can outperform at least one instance of MCMC, the goal is to clear the way for future research (which space constraints prohibit exploring extensively herein). That future research can then optimise the SMC Sampler and compare it with better-performing MCMC methods, such as TMCMC \cite{tmcmc} or HMC \cite{HMC}, in the context of both single and multiple chains (see above). In doing so, optimisations of SMC Samplers may include improved L-kernels, proposal distributions and a full comparison of resampling implementations (akin to that done in \cite{resampling_comparisons} in the context of a single core).
	
	The rest of the paper is organised as follows: in Section \ref{sec:dsm} we give some information about distributed memory architectures and MPI. In Section \ref{sec:smcmethods}, we describe Metropolis-Hastings and SMC methods with a focus on similarities and differences between Particle Filters and SMC Samplers. In Section \ref{sec:enhancements}, we introduce our novel implementation strategy. In Section \ref{sec:evaluation}, we describe and show the results of several exemplary case studies with a view to showing worst-case performance, maximum speed-up and space complexity of our MPI implementation of the SMC Sampler and its performance versus Metropolis-Hastings. In Section \ref{sec:conclusions}, we draw our conclusions and give suggestions for future improvements.
	
\section{Distributed Memory Model} \label{sec:dsm}
	Distributed memory architectures are a type of parallel system which are inherently different from shared memory architectures. In this environment, the memory is distributed over the cores and each core can only directly access its own private memory. Exchange of information stored in the memory of the other cores is achieved by sending/receiving explicit messages through a common communication network.
	
	The main advantages relative to shared memory architectures include scalable memory and computation capability with the number of cores and a guarantee of there being no interference when a core accesses its own memory.
	The main disadvantage is the cost of communication and consequent data movement. This may affect the speed-up relative to a single-core. 
	
	In order to implement the algorithms we discuss in this paper, we use Message Passing Interface (MPI) which is one of the most common programming models for distributed memory environments. In this model, the cores are uniquely identified by a rank, connected via communicators and they use Send/Receive communication routines to exchange messages.
	
\section{SMC and MCMC methods}\label{sec:smcmethods}
	In this section, we provide details about MCMC and SMC methods with a view to showing similarities and differences between Particle Filters and SMC Samplers. The reader is referred to \cite{delmoral}, \cite{Simon_1st} and \cite{Metropolis1953} for further details.

	\vspace{-5pt}
	\subsection{Sequential Monte Carlo methods}
		SMC methods apply the Importance Sampling principle to make Bayesian inferences. The main idea consists of generating $N$ statistically independent hypotheses called particles (or samples) at every given iteration $t$. The population of particles $\mathbf{x}_t \in \mathbb{Re}^{N \times M}$ is sampled from a user-defined proposal distribution $q(\mathbf{x}_t|\mathbf{x}_{t-1})$ such that $\mathbf{x}_t$ represents the pdf of the state of a dynamic model (in Particle Filters) or samples from a static target posterior distribution (in SMC Samplers\footnote{While it is not discussed here extensively, SMC Samplers can also be configured to offer improved performance in contexts where a Particle Filter struggles\cite{Maskellsmcs}.}). Each particle $\mathbf{x}_t^i$ is then assigned to an unnormalised importance weight $w_t^i = F(w_{t-1}^i, \mathbf{x}_t^i, \mathbf{x}_{t-1}^i)$, such that the array of weights $\mathbf{w}_t \in \mathbb{Re}^N$ provides information on which particle best describes the real state of interest. 

		The particles are however subjected to a phenomenon called degeneracy which (within a few iterations) makes all weights but one decrease towards $0$. This is because the variance of the weights is proven to increase at every iteration \cite{Simon_1st}. There exist different strategies to tackle degeneracy. The most common is to perform a resampling step which repopulates the particles by eliminating the most negligible ones and duplicating the most important ones. Different variants of resampling exist \cite{resampling_comparisons} and the chosen methodology is described in detail in Section \ref{subsec: key componets}. Resampling is only triggered when it is needed, more precisely when the (approximate) effective sample size
		\begin{align}
			\label{eq:neff}
			N_{eff} = \frac{1}{\sum_{i = 0}^{N - 1} (\mathbf{\tilde{w}}_t^i)^2}
		\end{align}  
		\noindent decreases below a certain threshold $N^*$ (which is usually set to $\frac{N}{2}$). $\mathbf{\tilde{w}}_t \in \mathbb{R}^{N}$ represents the array of the normalised weights, each of them calculated as follows:
		\begin{align}
			\label{eq:normalise}
			\tilde{w}_t^i  = \frac{w_t^i }{\sum_{j=0}^{N - 1} w_t^j}
		\end{align}  
		At every iteration, estimates are produced as a weighted sum of $\mathbf{x}_t$, weighted using $\mathbf{\tilde{w}}_t$.

		\subsubsection{Particle Filters}

			A range of different Particle Filter methods exist. This section provides a brief description of Sequential Importance Resampling (SIR), described by Algorithm \ref{alg:pf} in the appendix. 

			Let $\mathbf{X}_t \in \mathbb{Re}^M$ be the current state of the dynamic system that we want to estimate. At every time step $t$ a new measurement $\mathbf{Y}_t \in \mathbb{Re}^D$ is collected. In the SIR Filter, the weighted particles are initially drawn from the prior distribution $q(\mathbf{x}_0) = p(\mathbf{x}_0)$ and then drawn from the proposal distribution as follows: 
			\begin{align} \label{eq:pf_drawing_particles}
				\mathbf{x}_t^i \sim q(\mathbf{x}_t^i|\mathbf{x}_{t-1}^i, \mathbf{Y}_t)
			\end{align}
			The weights are initially set to $1/N$ and then computed as
			\begin{align}
				\label{eq:pf_drawing_weights}
				w_t^i = F(w_{t-1}^i, \mathbf{x}_t^i, \mathbf{x}_{t-1}^i) = w_{t-1}^i \frac{p(\mathbf{x}_t^i | \mathbf{x}_{t-1}^i) p(\mathbf{Y}_t|\mathbf{x}_t^i)}{q(\mathbf{x}_t^i|\mathbf{x}_{t-1}^i, \mathbf{Y}_t)}
			\end{align}
			\noindent The weights are then normalised and used to calculate $N_{eff}$ as in (\ref{eq:neff}). Then resampling is performed if needed. In the last step, the estimation of the state is given by the weighted mean of the particles.

		\subsubsection{SMC Samplers with recycling} \label{subsubsec: smcs}
			Like MCMC methods, the goal in the SMC Samplers is to draw samples from a static target distribution of interest $\pi_t(\mathbf{x}_t)$. The algorithm begins by drawing $N$ samples from the initial proposal $q(\mathbf{x}_0)$ and giving the $i$-th sample the weight $w_0^i = \frac{\pi_0(\mathbf{x}_0^i)}{q_0(\mathbf{x}_0^i)}$.

			After the first iteration, the samples are drawn from the forward Markov kernel, $q_t(\mathbf{x}_t|\mathbf{x}_{t-1})$, while the weights require backward Markov kernels $L_t(\mathbf{x}_{t-1}|\mathbf{x}_t)$ as follows: 		
			\begin{align}
				\label{eq:smcs_drawing_weights}
				w_t^i = F(w_{t-1}^i, \mathbf{x}_t^i, \mathbf{x}_{t-1}^i) = w_{t-1}^i\frac{\pi_t(\mathbf{x}_t^i)}{\pi_t(\mathbf{x}_{t-1}^i)}\frac{L_t(\mathbf{x}_{t-1}^i|\mathbf{x}_t^i)}{q_t(\mathbf{x}_t^i|\mathbf{x}_{t-1}^i)}
			\end{align}  
			As is the case for Particle Filters, after the importance weights evaluation and normalisation, the resampling step may be triggered depending on the value of $N_{eff}$. 

			In the vanilla SMC Sampler, estimates are performed according to the particles in the final iteration. The expected value is computed by multiplication of the particles at the final iteration $T$ with the corresponding weights. In~\cite{septier}, a novel recycling method is proposed. Instead of considering the particles from the last iteration as providing the outputs, estimates are computed using all particles from all iterations. Using the notation of this paper, estimates are performed as follows:
			\begin{align}
				\label{eq:recycling}
				\mathbf{\hat{f}} = \frac{\sum_{t=1}^T \mathbf{f}_t \tilde{c}_t}{\sum_{t=1}^T \tilde{c}_t}
			\end{align}  
			\noindent where $\mathbf{f}_t$ is calculated as
			\begin{align}
				\label{eq:estimation_at_t}
				\mathbf{f}_t = \sum_{i=0}^{N - 1} \mathbf{x}_t^i \tilde{w}_t^i
			\end{align}  
			\noindent and the normalisation constant\footnote{(\ref{eq: normconst}) is equivalent to (14) in \cite{septier}, albeit with simplified notation here.} is
			\begin{align} \label{eq: normconst}
				\tilde{c}_t = \int \pi (\mathbf{x}_t ) d\mathbf{x}_t \approx c_t = \frac{\sum_{i=0}^{N - 1} w_t^i}{\sum_{i=0}^{N - 1} w_{t-1}^i}
			\end{align}

			Algorithm~\ref{alg:smcsampler} in the appendix describes the SMC Sampler with the recycling method. 

	\vspace{-5pt}
	\subsection{Metropolis-Hastings Algorithm}
		The Metropolis-Hastings (MH) algorithm~(see Algorithm ~\ref{alg:MH} in the appendix) simulates a Markov chain where, at each iteration, a new sample, $\mathbf{x}^*$, is drawn from a proposal distribution. The new sample is accepted or rejected using the Rejection Sampling principle with acceptance probability $a = min\{1,\frac{\pi(\mathbf{x}^*) q(\mathbf{x}|\mathbf{x}^*)}{\pi(\mathbf{x})q(\mathbf{x}^*|\mathbf{x})}\}$. To reduce the dependency on the initial sample, the first (user-defined) $\tau$ samples are discharged (burn-in).

	\vspace{-5pt}
	\subsection{Key components of Particle Filters and SMC Samplers} \label{subsec: key componets}

		Algorithms \ref{alg:pf} and \ref{alg:smcsampler} in the appendix show that SIR Particle Filter and SMC Sampler with recycling share the same key components. 

		Importance Sampling is trivially parallelisable as (\ref{eq:pf_drawing_particles}), (\ref{eq:pf_drawing_weights}) and (\ref{eq:smcs_drawing_weights}) are element-wise operations. Hence, Importance Sampling achieves $O(1)$ time complexity for $P = N$ cores. 

		Expressions (\ref{eq:neff}), (\ref{eq:normalise}), (\ref{eq:estimation_at_t}), and the weighted mean of particles require Sum and then can be easily parallelised by using Reduction. The time complexity of any Reduction operation scales logarithmically with the number of cores. 

		Both algorithms invoke resampling if $N_{eff} < N^*$. Several alternative resampling steps have been proposed and a comparison between them is discussed in \cite{resampling_comparisons}. These algorithms solve the problem in $O(N)$ operations. The key idea of these algorithms is to process $\mathbf{w}_t$ to generate an array of integers called $\mathbf{ncopies} \in \mathbb{Z}^N$ whose $i$-th element, $ncopies^i$, indicates how many times the $i$-th particle has to be duplicated. It is easy to infer that $\mathbf{ncopies}$ has the following property:
		\begin{equation} \label{eq: ncopies}
			\sum\nolimits_{i=0}^{N-1} ncopies^i = N
		\end{equation}

		In previous work to parallelise Particle Filters described in \cite{Simon_1st}, \cite{Simon_2nd}, \cite{Lykourgos} and \cite{Alessandro}, Minimum Variance Resampling (MVR), a variant of Systematic Resampling in \cite{resampling_comparisons}, has always been the preferred resampling scheme. Since this paper is built on the results in \cite{Alessandro}, MVR will be the only variant of resampling we consider. This algorithm uses Cumulative Sum to calculate the CDF and then it generates $\mathbf{ncopies}$ such that the new population of particles has minimum ergodic variance. After that, it is necessary to perform a task called redistribute which duplicates $\mathbf{x}_t^i$ as many times as $ncopies^i$. This task has already been identified as bottleneck (see \cite{Simon_1st}, \cite{Simon_2nd}, \cite{Lykourgos}) and it will be discussed in detail in the next section. We note that the reset step (which sets all the weights to $1/N$) after redistribute is trivially parallelised.

		\subsubsection{Redistribute}

			The redistribute step is necessary to regenerate the population of particles and is a task which all resampling variants have in common. A naive and mature implementation can be found in \cite{Simon_1st}\cite{gordon}. The same is described by Algorithm \ref{alg: systematic_redistribute} in the appendix and referred to as Sequential Redistribute (S-R) in the rest of the paper. This routine simply iterates over $\mathbf{ncopies}$ and, for the $j-$th element, it copies $\mathbf{x}^j$ as many times as $ncopies^j$. Considering that $\mathbf{ncopies}$ follows (\ref{eq: ncopies}), it is easy to infer that S-R achieves $O(N)$ time complexity with a very low constant time. However, this algorithm is not trivial to parallelise because the workload among the processors cannot be readily distributed deterministically. This is because $ncopies^j$ could be equal to any value between $0$ and $N$. Parallelisation is even more complicated on distributed memory architectures since one core cannot directly access the memory of the other cores \cite{Alessandro}.

			In \cite{Simon_2nd}, it has been shown that, by using a top-down divide-and-conquer approach, redistribute can be parallelised. Starting from the root node, the key idea consists of sorting $\mathbf{ncopies}$ and moving the particles at every stage of a binary tree. This can be achieved by searching for a particular index called $pivot$ which perfectly divides the node into two balanced leaves. Once $pivot$ is identified, the node is split. In order to find $pivot$, Cumulative Sum (whose parallel implementation runs in $O(log_2N)$ steps \cite{Cumulative_Sum_1}) is performed and then $pivot$ is the first index where Cumulative Sum is equal to or greater than $\frac{N}{2}$. This routine is repeated recursively $log_2N$ times. Since Bitonic Sort is the chosen parallel sorting algorithm and it is known that its time complexity is equal to $O((log_2N)^2)$ with $P = N$ cores, then we can infer that this redistribute achieves $O((log_2N)^3)$ time complexity for the same level of parallelism. Sorting the particles is required to make sure that the splitting phase can be performed deterministically in $O(1)$.

			In \cite{Lykourgos}, the redistribute algorithm in \cite{Simon_2nd} was improved by making a subtle consideration: the workload can still be divided deterministically if we perform Bitonic Sort only once. After this single sort, the algorithm moves on to another top-down routine where we use rotational shifts to shift all particles on the left side of $pivot$ up to the left side of the node. This way the father node gets split into two balanced leaves. This algorithm is recursively performed $O(log_2N)$ times until the workload is equally distributed across the cores; then S-R is called. Algorithm \ref{alg: BSB-R} in the appendix summarises this routine and, in this paper, is described as Bitonic Sort Based Redistribute (B-R). Rotational shifts are faster than Bitonic Sort as the achieved time complexity is equal to $O(log_2N)$ and, therefore, the overall time complexity is improved to $O((log_2N)^2)$. In \cite{Lykourgos}, B-R has been implemented on MapReduce and, although it was significantly better than the algorithm in \cite{Simon_2nd}, its runtime for $512$ cores was up to $20$ times worse than a single-core S-R. 

			In \cite{Alessandro}, B-R has been ported to distributed memory architectures by using MPI and compared to a deterministic MPI implementation of S-R, in which one core gathers all particles from the other cores, performs S-R locally and scatters back the resulting array. To avoid misunderstanding, we refer to the MPI implementation of S-R in \cite{Alessandro} as Centralised Redistribute (C-R), which is described by Algorithm \ref{alg: centralised_redistribute} in the appendix. The results indicate that the scalability of the MPI implementation is improved relative to the scalability achieved using MapReduce because B-R on MPI could outperform C-R for at least $P = 64$ cores. Possible ways to improve B-R are discussed in the next session.

\section{Novel Implementation}\label{sec:enhancements}
	In this section, we consider ways to improve redistribute and how an MPI SMC Sampler could be an alternative to Metropolis-Hastings.

	\vspace{-5pt}
	\subsection{Improving single-core Bitonic Sort} \label{subsec: possible_ways}
		\begin{figure}[!htp]
			\centering
			\includegraphics[width=0.7\linewidth]{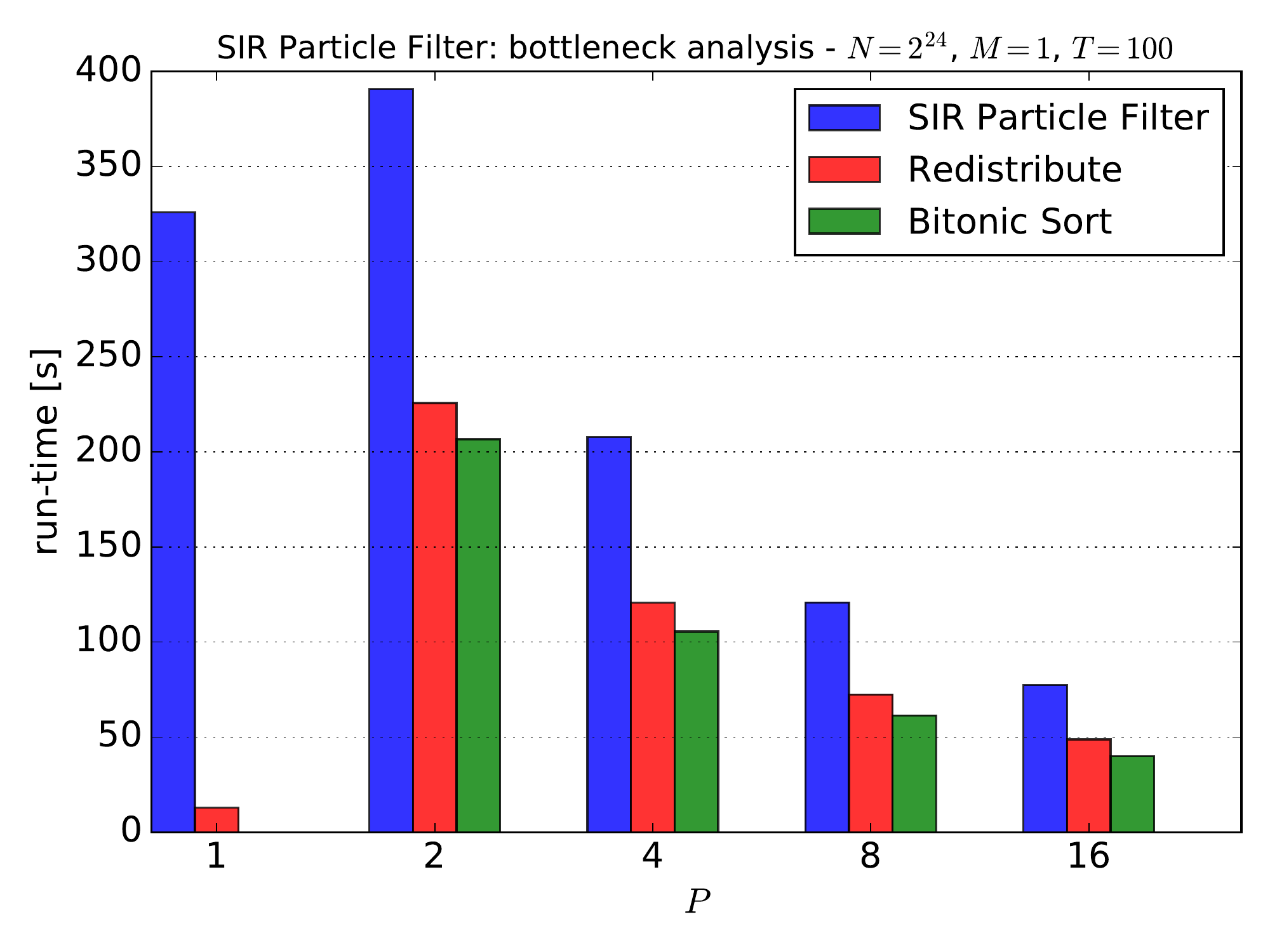}
			\caption{SIR Particle Filter: bottleneck analysis - $N = 2^{24}$, $M = 1$, $T = 100$}   
			\label{fig:bottleneck_analysis}
		\end{figure}

		As is observed here and has been discussed elsewhere in the literature \cite{Lykourgos}\cite{Alessandro}, the redistribute step is the bottleneck that complicates parallel implementation of Particle Filters. To ensure this is clear, we repeat an experiment from \cite{Alessandro} and report the results of the same SIR Particle Filter with B-R within using $N = 2^{24}$, $T = 100$ in Figure \ref{fig:bottleneck_analysis}. The run-times vs the number of cores $P$ for the entire Particle Filter, the constituent redistribute step and the subset of redistribute that is taken up with the Bitonic Sort step are given. As we can see, for $P > 1$, redistribute always accounts for at least $50\%$ of the total run-time and this proportion increases with $P$. For the same values of $P$, we can also observe that Bitonic Sort is by far the most computationally intensive task within redistribute and hence is the true bottleneck of the Particle Filter.
		\begin{figure}[!htp]
			\centering
			\includegraphics[width=0.7\linewidth]{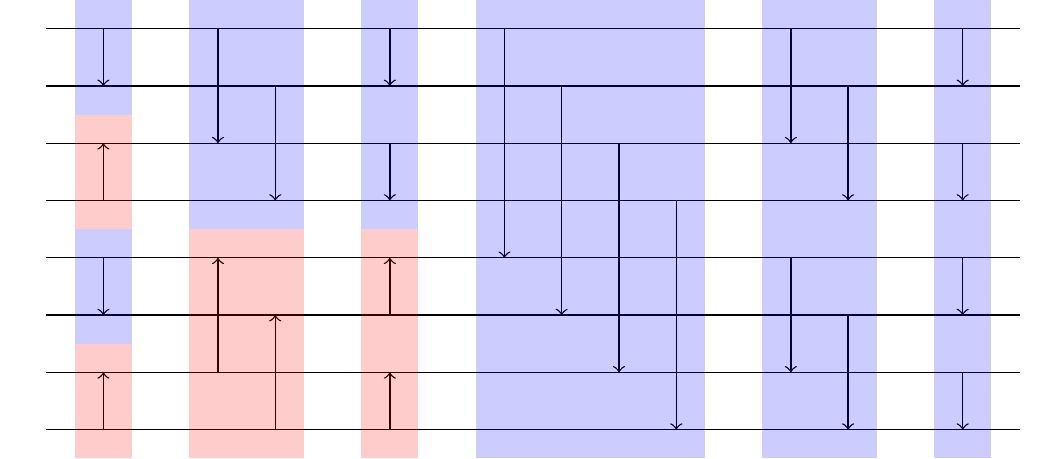}
			\caption{Bitonic Sort \& Nearly Sort - Sorting Network}   
			\label{fig:sorting network}
		\end{figure}

		Bitonic Sort is a very fast comparison-based parallel sorting algorithm \cite{Batcher68}. This algorithm uses a divide-and-conquer approach to first divide the input sequence into a series of Bitonic sequences\footnote{A Bitonic sequence is a sequence of $N$ keys in which the first $N/2$ keys are sorted in increasing order, while the last $N/2$ keys are sorted in decreasing order.}. Then the Bitonic sequences are recursively merged together until the algorithm returns a single monotonic sorted sequence. A possible sorting network which can be used is illustrated in Figure \ref{fig:sorting network}. Each horizontal wire represents a key, the vertical arrows connect the input keys for a comparison and the direction represents the order of the keys after the comparison has occurred. The coloured blocks represent application of Bitonic Merge (blue or red if Merge is called in increasing or decreasing order respectively).

		It has been proven that, given a generic array of $N$ elements, Bitonic Sort solves the problem in $O(N(log_2N)^2)$ comparisons \cite{Batcher68}. Bitonic Sort is, however, suitable to run in parallel by making $P$ cores work on different chunks of the input array. In this case, each wire (or groups of wires) in Figure \ref{fig:sorting network} may also represent a core (or the elements that each core operates on). When $P = N$, it is easy to infer that the achieved time complexity is equal to $O((log_2N)^2)$. More generically, we can say that for any number of cores $P \leq N$ the time complexity is equal to
		\begin{equation} \label{eq: Bitonic_Sort_time_complexity}
			O\left(\frac{N}{P} \left( log_2\left(\frac{N}{P}\right) \right)^2 + \frac{N}{P}\left( log_2P \right)^2\right)
		\end{equation}
		$\frac{N}{P}(log_2(\frac{N}{P}))^2$ is the time complexity to perform Bitonic Sort locally before the cores start interacting with each other. This term is definitely dominant, especially for low values of $P$. One possible way to improve Bitonic Sort (and by extension redistribute) is to substitute the serial Bitonic Sort algorithm with a better single-core sorting algorithm, as was been suggested in \cite{Alessandro}.

		In the literature, there are plenty of alternatives to Bitonic Sort available. Algorithms such as Quicksort, Mergesort and Heapsort, for example, achieve $O(Nlog_2N)$ time complexity. Quicksort is on average faster than Mergesort and Heapsort. However, Quicksort's choice of its pivot can severely influence the performance: it is known, in fact, that Quicksort's worst-case time complexity is $O(N^2)$. This occurs when the  pivot chosen at every iteration is equal to either the minimum or the maximum of the available keys. Although this case is statistically very rare in several modern applications, in the case of SMC methods the worst-case scenario is however often encountered: $\mathbf{ncopies}$ has to be sorted and, since (\ref{eq: ncopies}) holds, there is a high probability that $0$ is picked as Quicksort's pivot, i.e. a high probability that the pivot is the minimum element. 

		Heapsort achieves $O(Nlog_2N)$ time complexity in all cases except when all keys are equal. In this special although rather unlikely case, the time complexity is $O(N)$. However, Mergesort is perfectly deterministic and data-independent and represents a valid alternative to Bitonic Sort which we consider in the experiment in Section \ref{subsec: bottleneck}. A Bitonic Sorter with Mergesort performed locally achieves the following time complexity:
		\begin{equation} \label{eq: Bitonic_Sort_with_MS_time_complexity}
			O\left(\frac{N}{P}log_2\left(\frac{N}{P}\right) + \frac{N}{P}(log_2P)^2\right)
		\end{equation}

		We also observe that $\mathbf{ncopies}$ is an array of integers. Hence, one could locally use linear time sorting algorithms such as Counting Sort or Radix Sort (which are both only applicable to arrays of integers). Although Counting Sort has deterministic and data-independent time complexity, its space complexity is data-dependent. This is because Counting Sort allocates a temporary array with as many elements as $max - min + 1$. In the worst-case $max = N$, $min = 0$ and since $N$ could be very high, the temporary array may not fit within the local memory of a single machine. This problem is shared with C-R and the impact of this issue will be discussed in Section \ref{subsec: unfeasibility}. On the other hand, Radix Sort is a feasible deterministic solution. However, Radix Sort is data-dependent because its time complexity is actually $O(C \cdot N)$ where the constant $C$ is equal to the number of digits of the maximum element (which can be $N$ in the worst-case). Therefore, Radix Sort may be too slow when $N$ is high and its run-time may fluctuate too much as a function of the input. 

		In summary, we are looking for a parallel sorting algorithm that works with integer numbers and is deterministic and data-independent with respect to both time and space complexity. While a combination of Bitonic Sort and Mergesort within a core achieves these aims, in the next two sessions, we go on to develop an improved strategy that is sufficient for our needs and does not require sort at all. 

	\vspace{-5pt}	
	\subsection{Nearly Sort: an alternative to single-core sorting}
		In \cite{Simon_1st}, sorting was used extensively. In B-R, rotational shifts are used $log_2P$ times while Bitonic Sort is used only once. This replacement of sort with rotational shift has improved the time complexity (from $O((log_2N)^3)$ to $O((log_2N)^2))$. However, it has also led to a more subtle consideration: by observing the input of rotational shifts we can infer that we do not actually need to perfectly sort the particles to divide the workload deterministically. This condition is always satisfied as long as stage by stage the particles that have to be duplicated are separated from those that do not. To make things more clear we first provide the following definition.

		\begin{Def}
			Let $\mathbf{g}$ be a sequence of $N$ elements. $\mathbf{g}$ is called Nearly-Sorted sequence when it has the following shape:
			\begin{math}
				\left[0, ..., 0, g^0, g^1, ..., g^{m-1}\right]
			\end{math}
			where $g^i > 0$ $\forall i = 0, 1, ..., m-1$ and $0 \leq m \leq N$. $\mathbf{g}$ is an ascending Nearly-Sorted sequence if the first elements of $\mathbf{g}$ are $0$ and a descending Nearly-Sorted sequence if the final elements are $0$.
		\end{Def}

		We can infer that the workload can be divided deterministically if $\mathbf{ncopies}$ is a Nearly-Sorted sequence. In B-R, this condition is ensured by sorting before the subsequent parts of the redistribute step. While there are single-core sorting algorithms that achieve $O(N)$ time complexity, these algorithms do not satisfy our need for deterministic run-time and storage. However, it is possible to use a single core nearly-sort for an array of integers with a deterministic and data-independent approach with $O(N)$ time complexity.

		Algorithm \ref*{alg: Serial_Nearly_Sort} in the appendix, which we have called Sequential Nearly Sort (S-NS), declares two iterators $l$ and $r$ which respectively point at the first and the last element of $\mathbf{ncopies}$. Step by step, the $i$-th element of $\mathbf{ncopies}$ is considered and if the value is higher than $0$ then the particle is copied to the right end of the output array. If not, it gets copied to the left end. The output $\mathbf{ncopies}_{new}$ will then be an ascending Nearly-Sorted sequence. S-NS requires $N$ iterations of the for loop, which means that it achieves $O(N)$ time complexity or $O(\frac{N}{P})$ if we consider that each core owns $\frac{N}{P}$ elements. 

		S-NS is, therefore, a very good alternative to Serial Bitonic Sort, Mergesort, Heapsort and Radix Sort. This is because it achieves low time complexity with deterministic and data-independent run-time and space complexity. 

	\vspace{-5pt}
	\subsection{Parallel Nearly Sort}	
		We want S-NS to be used as part of a parallel algorithm which generates a Nearly-Sorted sequence from a random input one. We now discuss how to achieve this.

		\begin{Def}
			Let $\mathbf{h}$ be a sequence of $N$ elements. $\mathbf{h}$ is called a Nearly-Bitonic sequence when it is possible to find an index $k$ which splits $\mathbf{h}$ into two monotonic Nearly-Sorted sequences.
		\end{Def}

		One could use S-NS and the same sorting network of Bitonic Sort to first divide the input into a series of Nearly-Bitonic sequences and then to recursively merge the sequences together until we generate a monotonic Nearly-Sorted sequence at the last step. 

		We need to adapt Bitonic Merge such that it processes a Nearly-Bitonic sequence and returns a monotonic Nearly-Sorted sequence. We call this algorithm Nearly Merge. Stage-by-stage, one core with MPI rank $i$ is coupled with another core with MPI rank $j$. The assumption is that each core owns a Nearly-Sorted sequence of keys such that the combination of both is necessarily a Nearly-Bitonic sequence. Stage by stage, the cores call MPI\textunderscore Sendrecv to exchange their local data. Then they consume a complementary subset of $\frac{N}{P}$ elements. Depending on the direction of the arrow in the sorting network (see again Figure \ref{fig:sorting network}), one core will start consuming the $0$s first and then the positive elements while the other core will do the opposite. This way, the $0$s will be confined to one end of the output array separated from the positive elements. 	
		\begin{figure}[!htp]
			\centering
			\includegraphics[width=0.9\linewidth]{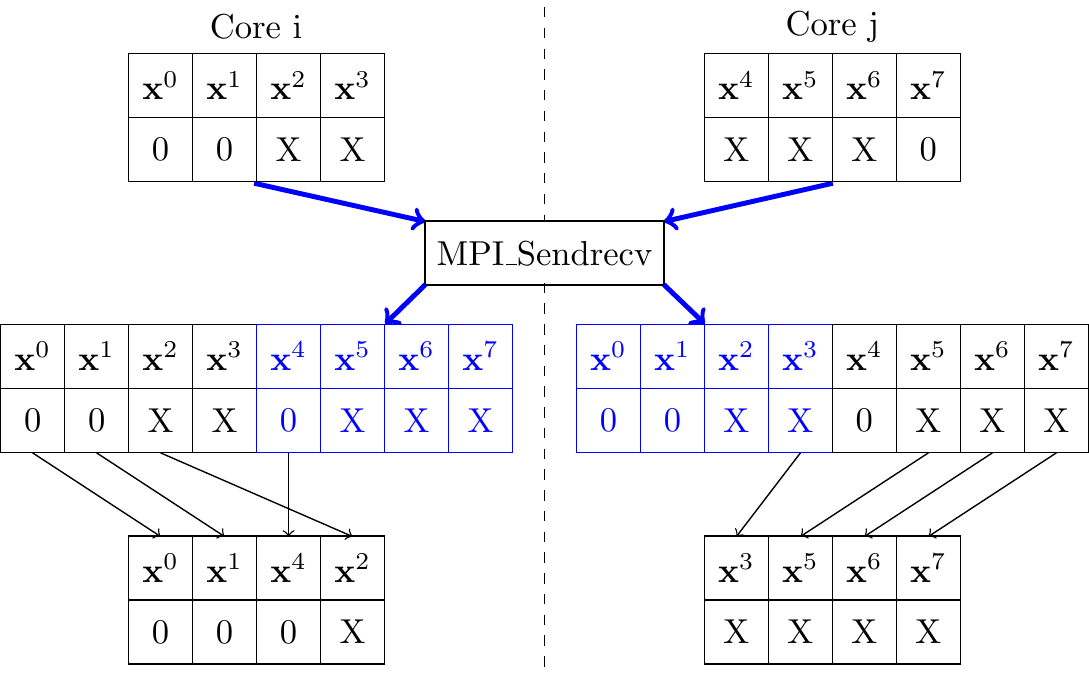}
			\caption{Nearly Merge - example}   
			\label{fig:nearly merge}
		\end{figure}

		Figure \ref{fig:nearly merge} illustrates a possible example of Nearly Merge where each core owns $4$ keys; the positive elements are padded with Xs for brevity. By extension, each core owns exactly $\frac{N}{P}$ particles and performs the same amount of writes to memory. Therefore, Nearly Merge achieves $O(\frac{N}{P})$ time complexity just as S-NS does. We can infer that the overall time complexity for Nearly Sort is equal to 
		\begin{equation} \label{eq: NS_time_complexity}
			O\left( \frac{N}{P} + \frac{N}{P}(log_2P)^2\right) 
		\end{equation}

\begin{figure}[!htp]
			\centering
			\includegraphics[width=1\linewidth]{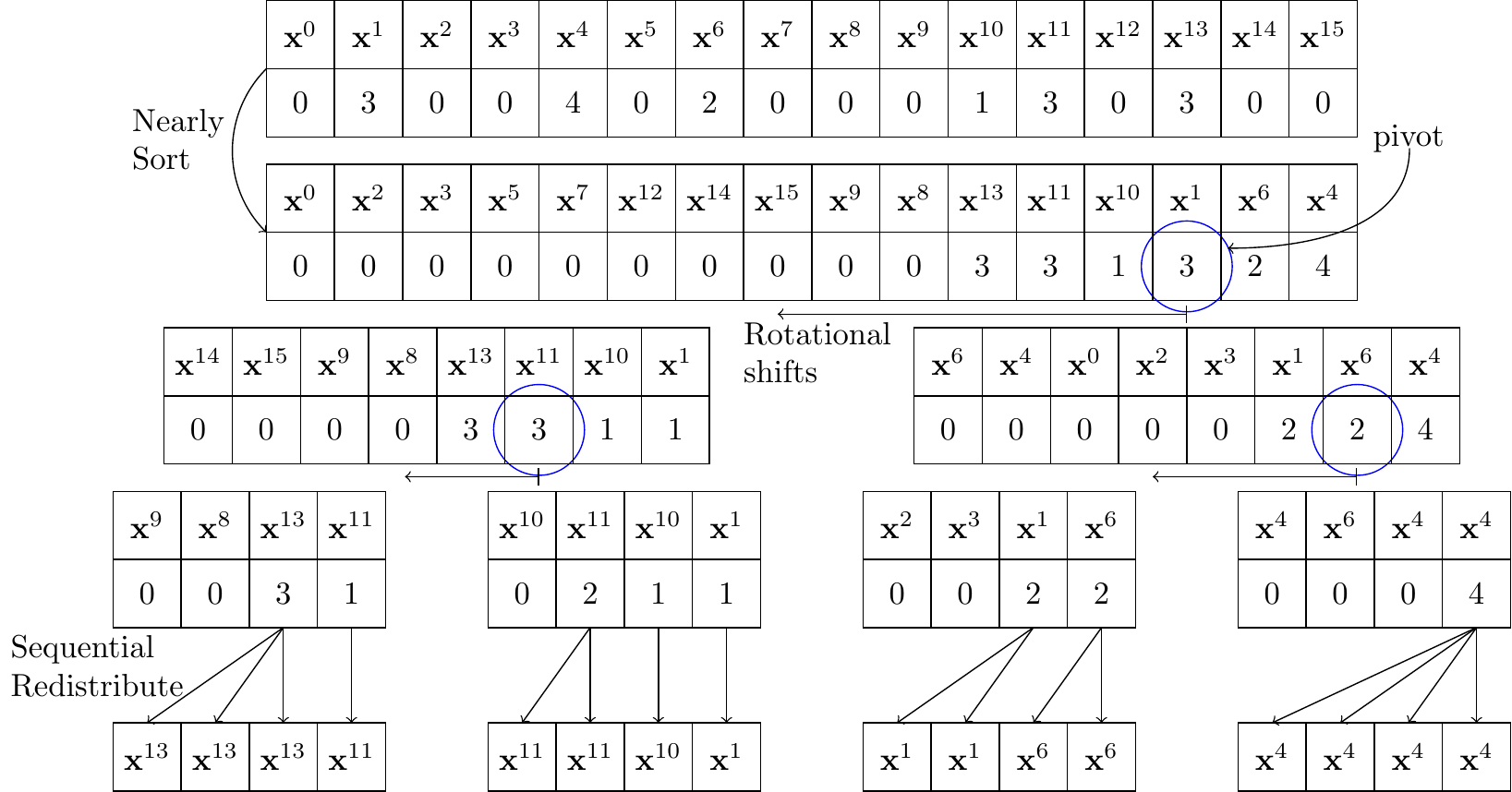}
			\caption{Nearly Sort Based Redistribute}   
			\label{fig:nearly redistribute}
		\end{figure}

		This algorithm has asymptotically the same time complexity of Bitonic Sort when $P = N$, but the time complexity for the serial algorithm is improved by a factor of $O((log_2(\frac{N}{P}))^2)$. Therefore, we expect this algorithm to outperform both Bitonic Sort and a Bitonic Sorter with Mergesort performed locally. By extension, if we exchange Bitonic Sort with Nearly Sort in B-R we expect to have better performance. Algorithm \ref{alg: NSB-R} in the appendix describes the new routine. A possible example for $N = 16$ and $P = 4$ is shown in Figure \ref{fig:nearly redistribute}. From now on, we refer to this algorithm as Nearly Sort Based Redistribute (N-R).

		In SMC methods $\mathbf{ncopies}$ is the array to (nearly) sort and each key $ncopies^i$ is necessarily coupled to the particle $\mathbf{x}^i \in \mathbb{Re}^M$. (\ref{eq: NS_time_complexity}) must then be extended to:
		\begin{equation} \label{eq: multi-state Nearly Sort time complexity}
			O\left(\frac{M \cdot N}{P} + \frac{M \cdot N}{P}(log_2P)^2\right)
		\end{equation}

		We denote that (\ref{eq: multi-state Nearly Sort time complexity}) can qualitatively describe the time complexity of N-R and, by extension, the time complexity of an SMC method which uses N-R within. The same conclusions about (\ref{eq: NS_time_complexity}) and (\ref{eq: multi-state Nearly Sort time complexity}) can be made about (\ref{eq: Bitonic_Sort_time_complexity}) and (\ref{eq: Bitonic_Sort_with_MS_time_complexity}) but they are left out for brevity.

	\vspace{-10pt}	
	\subsection{Single SMC Sampler vs Single Metropolis-Hastings} \label{subsec: smsc_vs_mh}
		Metropolis-Hastings and the SMC Sampler perform sampling from a target distribution and they both provide an accurate result for a sufficiently large number of iterations. However, the details of the two approaches differ substantially. As we have discussed in Section \ref{sec:smcmethods}, Metropolis-Hastings uses a Markov Chain to generate each sample one by one based on the history of the previous samples. This approach makes a single instance of Metropolis-Hastings hard to parallelise in a problem-agnostic way. On the other hand, the SMC Sampler is a population-based algorithm where all samples are processed independently and concurrently during each iteration. 

		Now let the total simulation-time for each algorithm be fixed to $\Delta$ seconds. After $\Delta$ seconds, Metropolis-Hastings and the SMC Sampler will have performed $T_{MH}$ and $T_{SMC}$ iterations respectively and provide a solution with a certain root mean squared error. Since a single Metropolis-Hastings is hard to parallelise, we cannot increase its accuracy without running the simulation for longer than $\Delta$ seconds. However, a single SMC Sampler can improve its throughput or accuracy by taking advantage of its inherent parallelism. In an ideal world, $2$ cores can perform $T_{SMC}$ iterations in $\frac{\Delta}{2}$ seconds, but they can also, and most importantly, run $2T_{SMC}$ iterations in $\Delta$ seconds. This means they can achieve better accuracy with the same run-time. By extension, $P$ cores can ideally run $P \cdot T_{SMC}$ iterations in $\Delta$ seconds but they will achieve a much better accuracy than a single core is capable of.

		The main goal of this paper is to prove that a $P$ core MPI SMC Sampler can be more accurate over the same run-time than Metropolis-Hastings when they sample from the same target distribution and use the same proposal. A more exhaustive explanation with experimental results is provided in Section \ref{subsubsec: smcs_vs_mcmc_description}.
		
\section{Case Studies and results}\label{sec:evaluation}
	In this section, we briefly describe the experiments we make and we analyse the results. Table \ref{tab:sysdetails} provides details about Barkla and Chadwick, the two platforms we use for the described experiments. Barkla is the preferred cluster for the majority of the experiments as it can provide more resources.

	\vspace{-5pt}
	\subsection{Bottleneck} \label{subsec: bottleneck}
		To evaluate the improvements in the bottleneck, we first focus on the sorting phase. Then we compare N-R, B-R and C-R. $M = 1$ in this first experiment for brevity.
		\begin{figure}[ht!]
			\centering
			\begin{subfigure}[b]{0.375\textwidth}
				\includegraphics[width=\columnwidth]{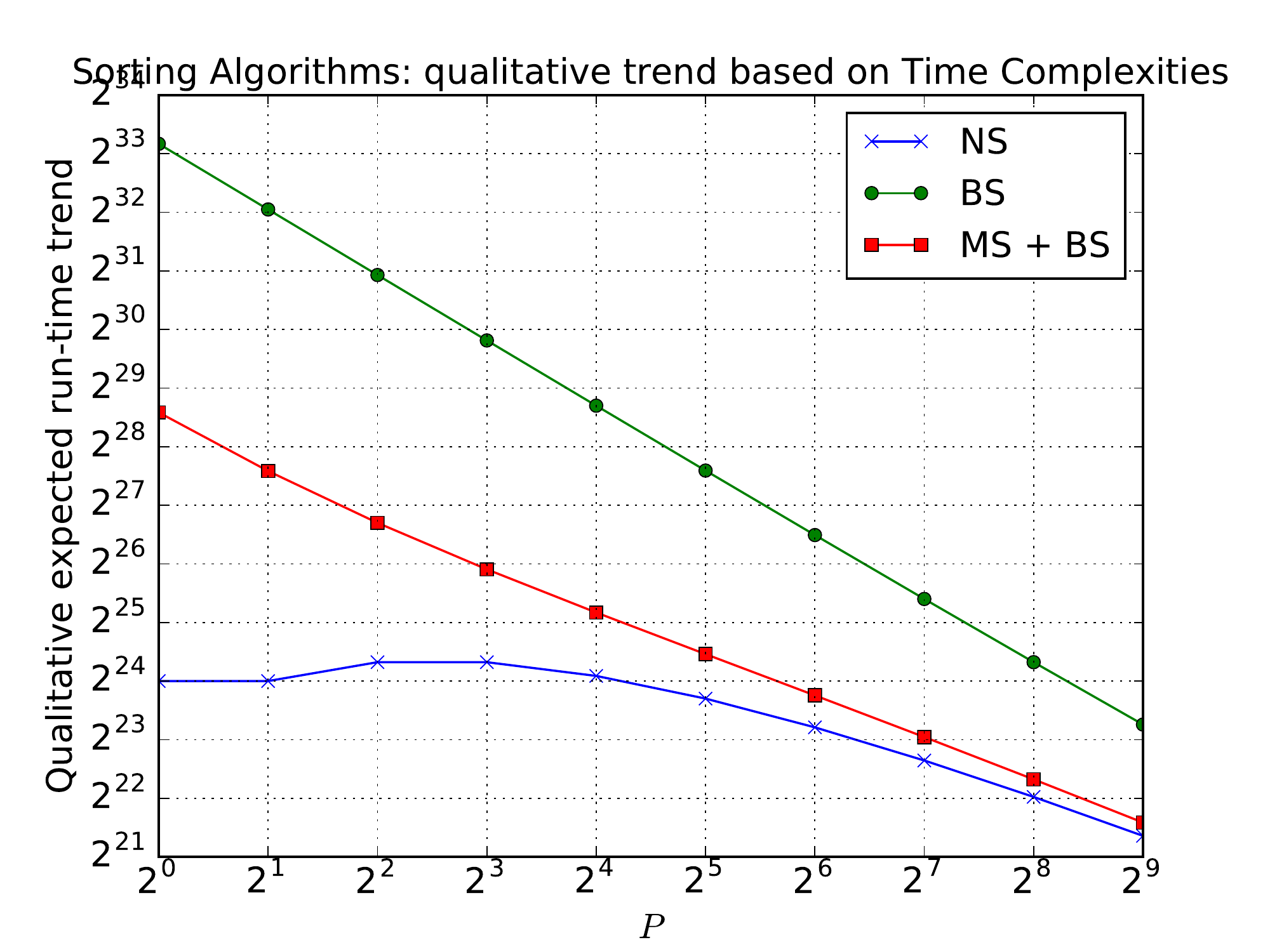}
				\caption[]%
				{{Sorting}}    
				\label{fig:sort_time_complexities_theory}
			\end{subfigure}
			\begin{subfigure}[b]{0.375\textwidth}    
				\includegraphics[width=\columnwidth]{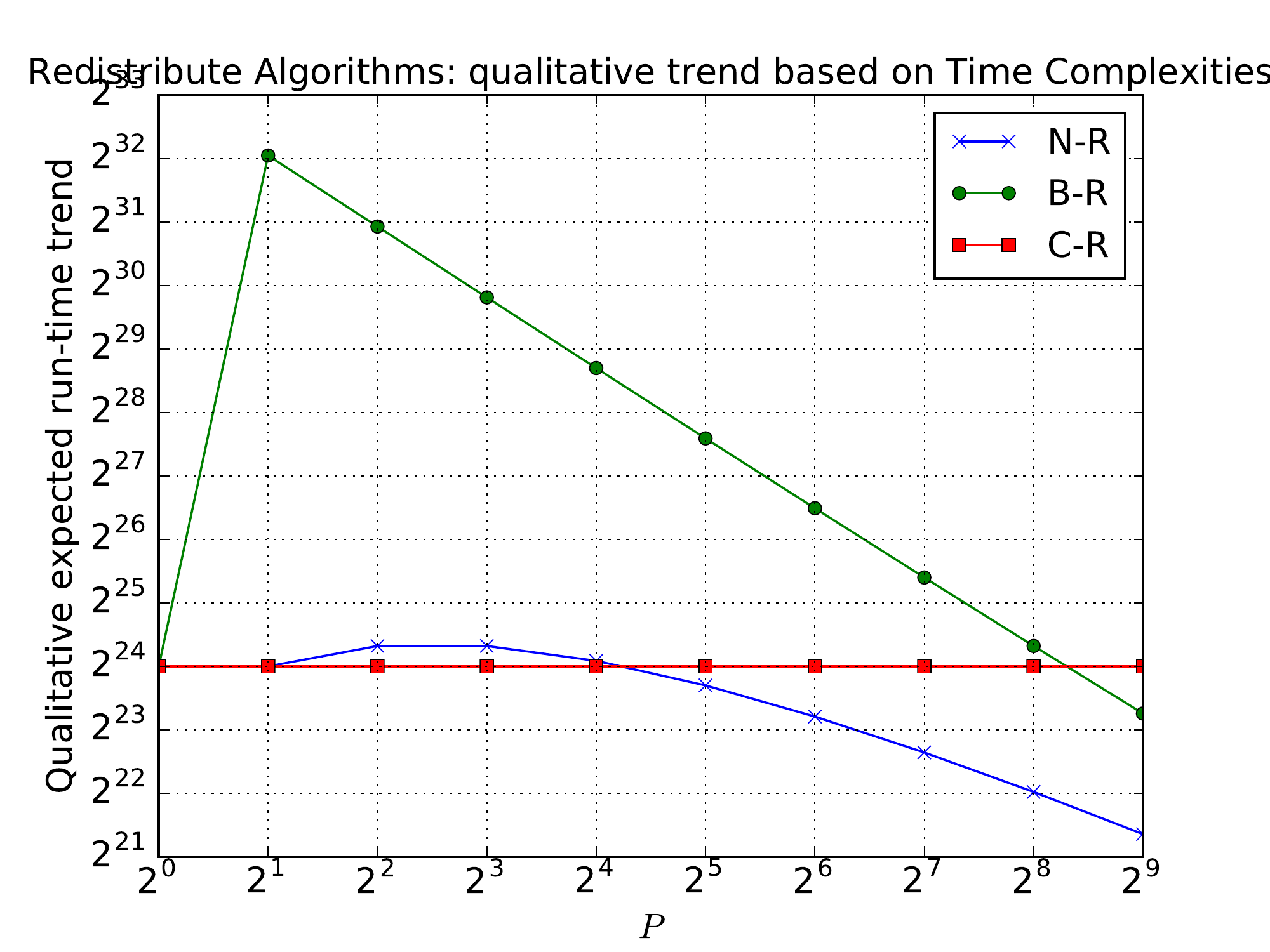}
				\caption[]%
				{{Redistribute}}    
				\label{fig:redistribute_time_complexities_theory}
			\end{subfigure}
			\caption{Bottleneck: theoretical run-time trend} \label{fig:Bottleneck_theoretical_runtime_tred}
		\end{figure}

		\subsubsection{Sorting} \label{subsubsec: eval_NS_vs_BS}
			As we outlined in Section \ref{sec:enhancements}, Bitonic Sort is the slowest task in B-R. In this experiment, we compare three different deterministic sorting/nearly sorting algorithms: Bitonic Sort (BS), Bitonic Sort with Mergesort performed locally (BS+MS) and Nearly Sort (NS). These algorithms are compared by passing in input the same two arrays: $\mathbf{ncopies}$ and $\mathbf{x}$. $\mathbf{ncopies}$ represents the array of the numbers of copies and hence it is an array of integers. It is generated randomly according to (\ref{eq: ncopies}) by using a Gaussian random generator followed by MVR. $\mathbf{x}$ represents an array of single-dimension particles and hence it is an array of floating point numbers which are generated by a Gaussian random generator. 

			The experiments have been run on Barkla for $N = 2^{10}, 2^{17}, 2^{24}, 2^{31}$ particles and increasing numbers of cores $P = 1, 2, 4, 8, ..., 512$. Both $N$ and $P$ must necessarily be equal to power of $2$ numbers, due to the constraint of Bitonic Sort and Nearly Sort. Each experiment has been run $20$ times and we report the median of the sampled run-times vs the number of cores.

			As we can see in Figure \ref{fig:bottleneck_results}, NS does not scale for up to $8$ cores. This trend might seem odd but it can be explained by analysing the time complexity of NS described by (\ref{eq: NS_time_complexity}). Figure \ref{fig:sort_time_complexities_theory} describes the qualitative trend of (\ref{eq: Bitonic_Sort_time_complexity}), (\ref{eq: Bitonic_Sort_with_MS_time_complexity}) and (\ref{eq: NS_time_complexity}). When $P = 1$, the quasilinear term in (\ref{eq: NS_time_complexity}) is equal to $0$. However, for $2 \leq P \leq 8$, the quasilinear term offsets the improvement associated with the linear term. Theoretically, the run-time should have positive speed-ups for $P = 32$. In the measured values for $N \geq 2^{17}$, this happens when $P \geq 32$ or $P \geq 64$ cores, depending on $N$. We associate this slight discrepancy to the additional communication cost associated with larger numbers of particles.

			However, the most important result is that NS is significantly faster than the other algorithms and especially BS for a low number of cores. Then, when $P$ increases the performance of both algorithms become closer because the time complexity of both algorithms is asymptotically equal to $O((log_2N)^2)$, as underlined in Section \ref{sec:enhancements}. These results suggest that using Bitonic Sort or Nearly Sort results in similar run-time for $P \geq 512$ but, using Nearly Sort may lead to significant improvements for $P < 512$. This means that the crossing point with respect to the run-time of C-R may be shifted to the left side of the graph, relative to the results in \cite{Alessandro}.

			The results for $N = 2^{10}$ keys show that BS and BS+MS stop scaling for a very low value of $P$. The reasons behind this result have required further investigation. For a very low number of keys, the granularity of the pipeline is already fine. In other words, the computation cost is already comparable with the communication cost and using more cores does not provide any scalability. NS is also affected by the same problem. In addition, the time complexity of Nearly Sort is necessarily higher than $O(N)$ for $P \leq 8$ cores. For these two reasons NS always returns negative speed-ups.

		\subsubsection{B-R vs N-R vs C-R} \label{subsubsec: eval_redistribute}
\begin{figure*}[!htp]
				\centering
				\begin{subfigure}[b]{0.33\textwidth}
					\centering
					\includegraphics[width=1\linewidth]{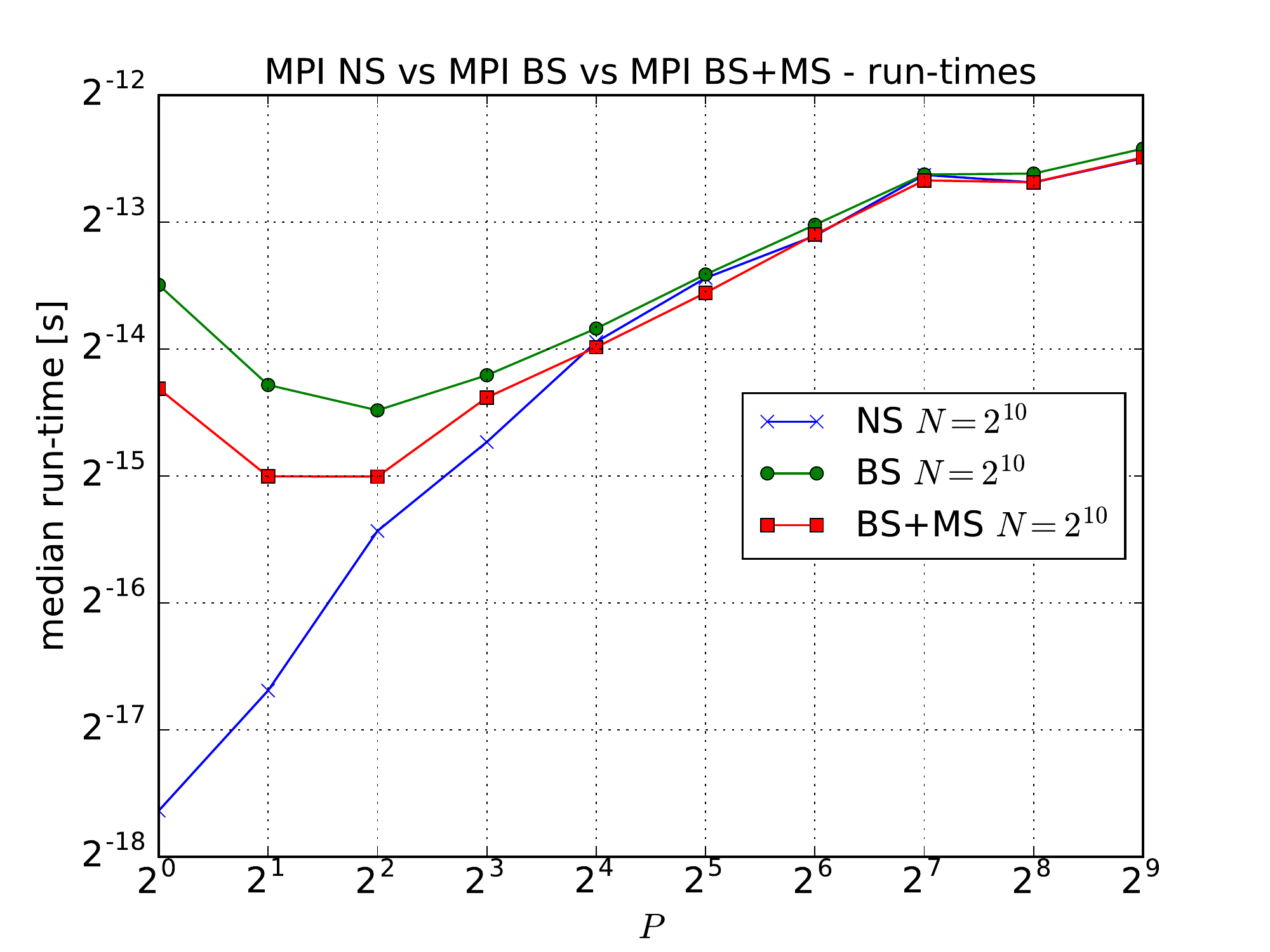}
					\caption[]%
					{{\small NS, BS and BS+MS for $N = 2^{10}$}}    
					\label{fig:ns_vs_bs_runtime_10}
				\end{subfigure}
				\hspace{1 cm}
				\begin{subfigure}[b]{0.33\textwidth}  
					\centering 
					\includegraphics[width=1\linewidth]{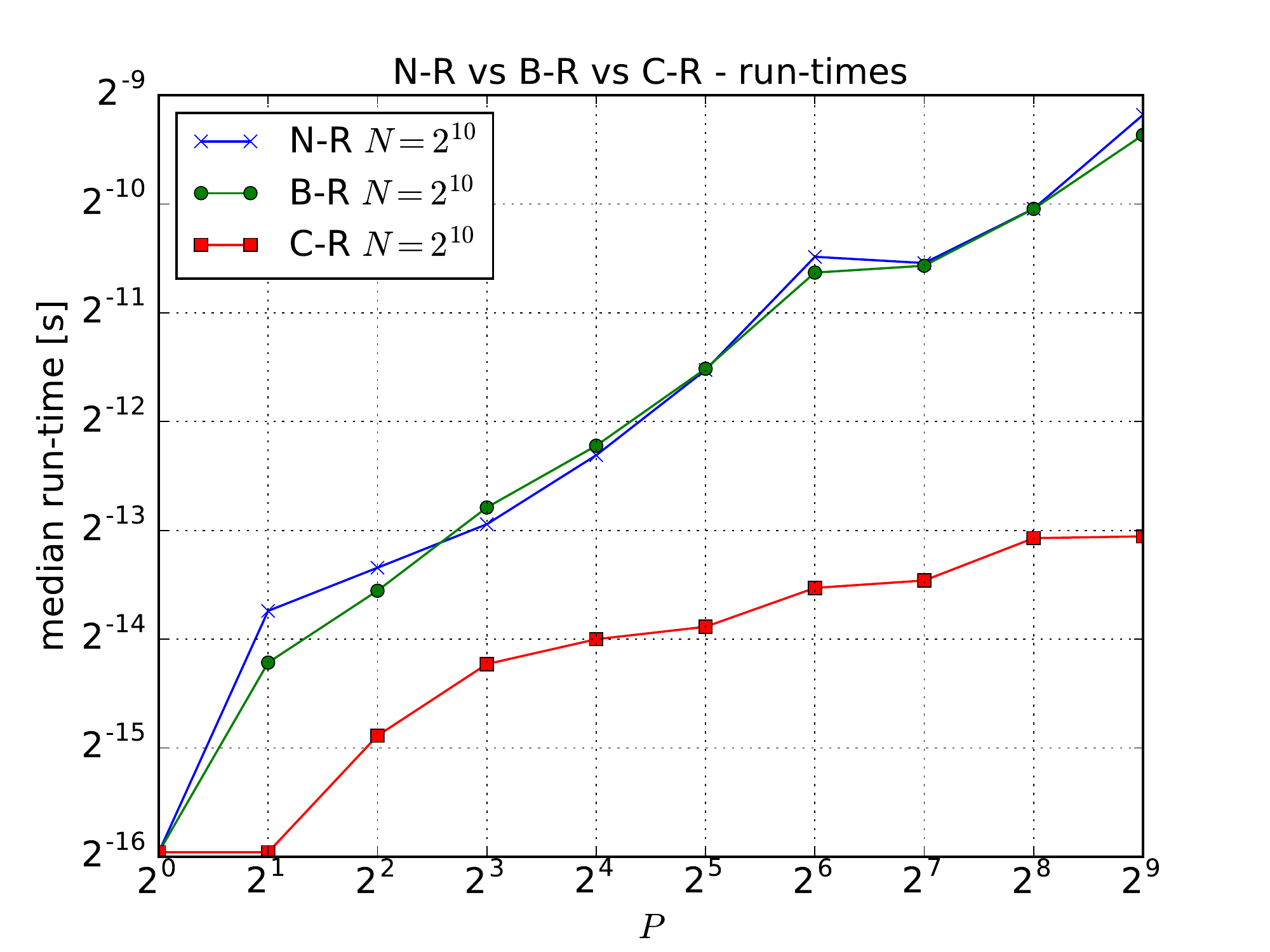}
					\caption[]%
					{{\small N-R, B-R and C-R for $N = 2^{10}$}}    
					\label{fig:redistribute_runtime_10}
				\end{subfigure}
				\vskip\baselineskip
	
				\centering
				\begin{subfigure}[b]{0.33\textwidth}
					\centering
					\includegraphics[width=1\linewidth]{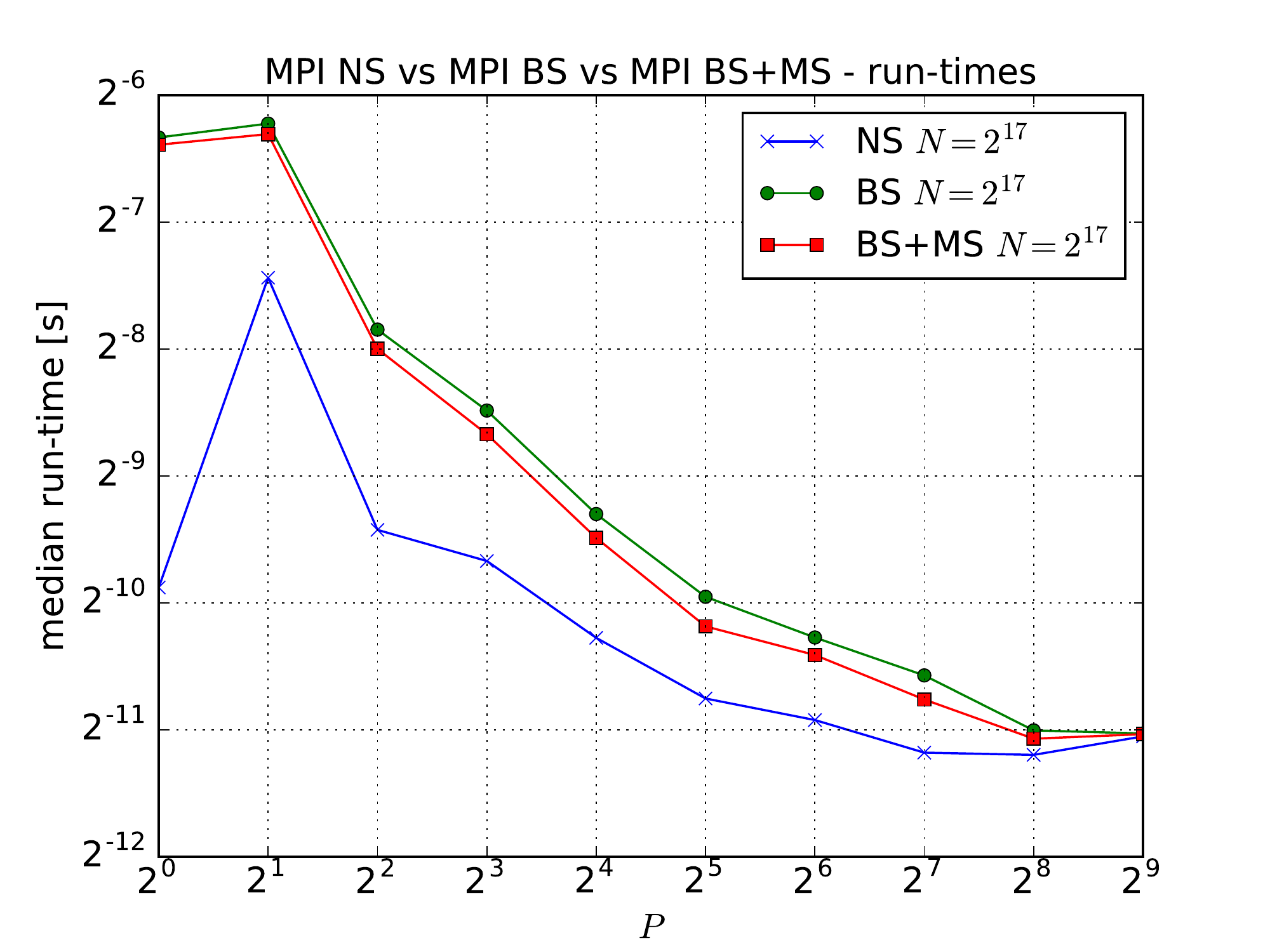}
					\caption[]%
					{{\small NS, BS and BS+MS for $N = 2^{17}$}}    
					\label{fig:ns_vs_bs_runtime_17}
				\end{subfigure}
				\hspace{1 cm}
				\begin{subfigure}[b]{0.33\textwidth}  
					\centering 
					\includegraphics[width=1\linewidth]{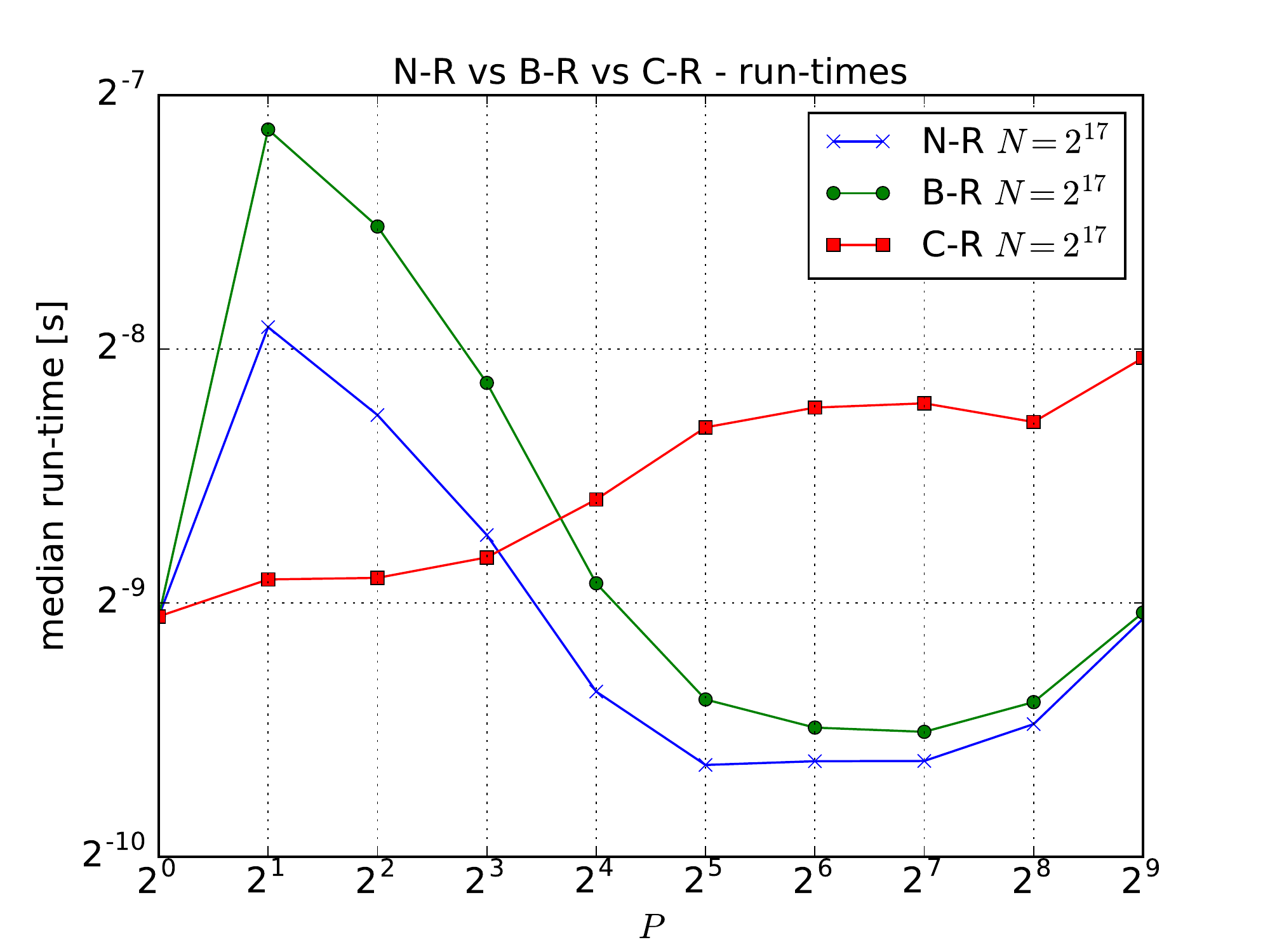}
					\caption[]%
					{{\small N-R, B-R and C-R for $N = 2^{17}$}}    
					\label{fig:redistribute_runtime_17}
				\end{subfigure}
				\vskip\baselineskip
	
				\centering
				\begin{subfigure}[b]{0.33\textwidth}
					\centering
					\includegraphics[width=1\linewidth]{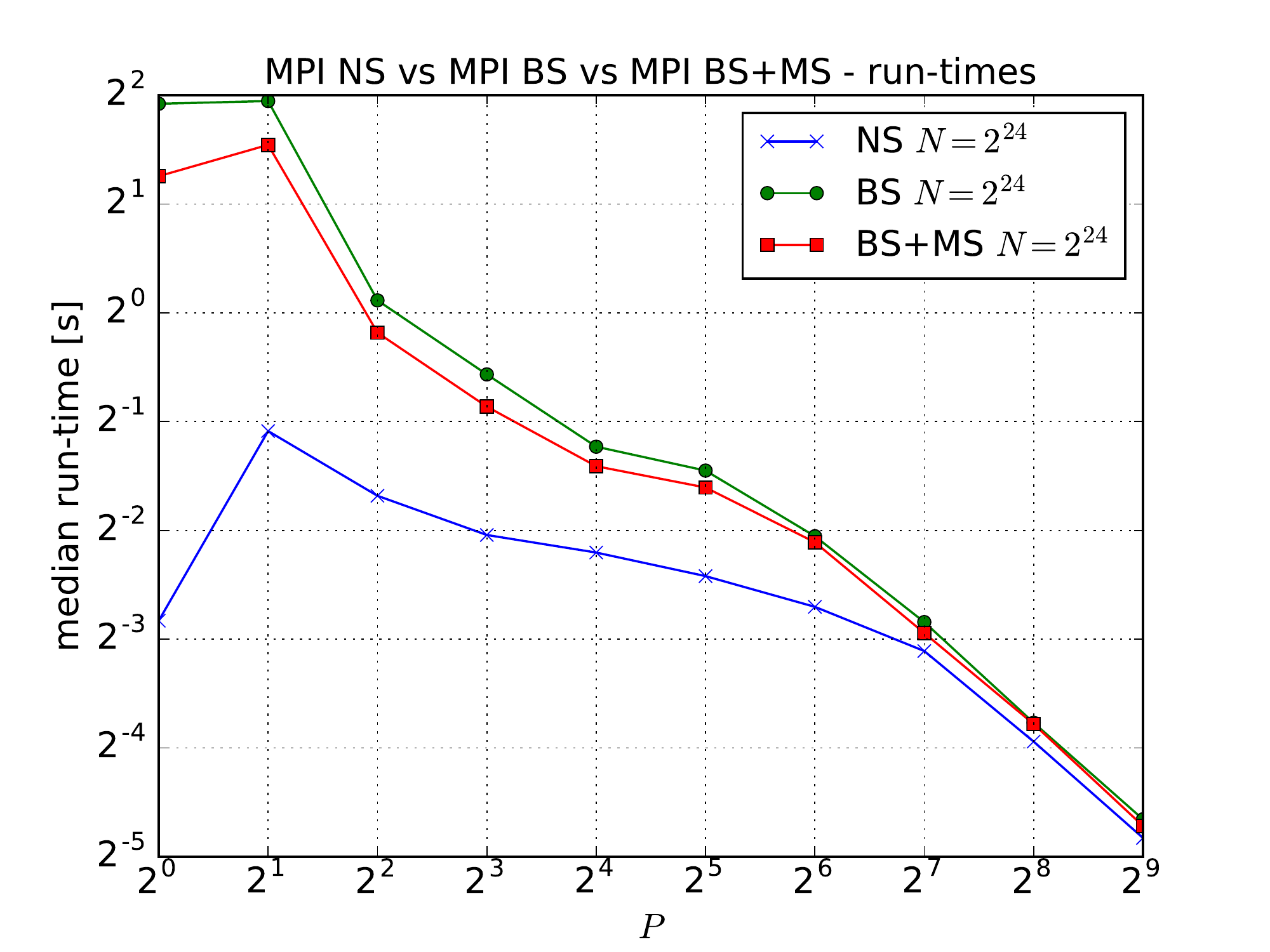}
					\caption[]%
					{{\small NS, BS and BS+MS for $N = 2^{24}$}}    
					\label{fig:ns_vs_bs_runtime_24}
				\end{subfigure}
				\hspace{1 cm}
				\begin{subfigure}[b]{0.33\textwidth}  
					\centering 
					\includegraphics[width=1\linewidth]{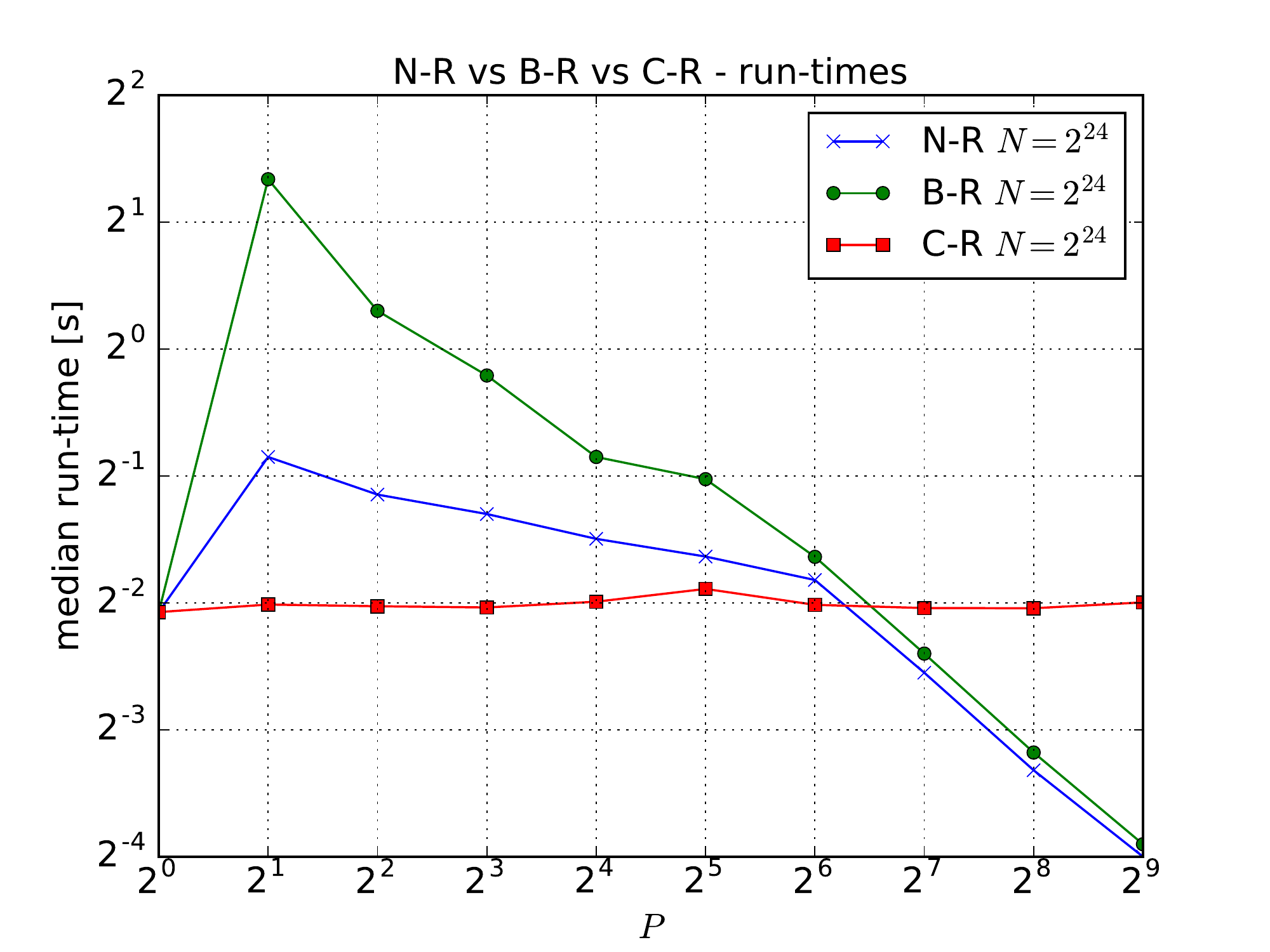}
					\caption[]%
					{{\small N-R, B-R and C-R for $N = 2^{24}$}}    
					\label{fig:redistribute_runtime_24}
				\end{subfigure}
				\vskip\baselineskip
	
				\centering
				\begin{subfigure}[b]{0.33\textwidth}
					\centering
					\includegraphics[width=1\linewidth]{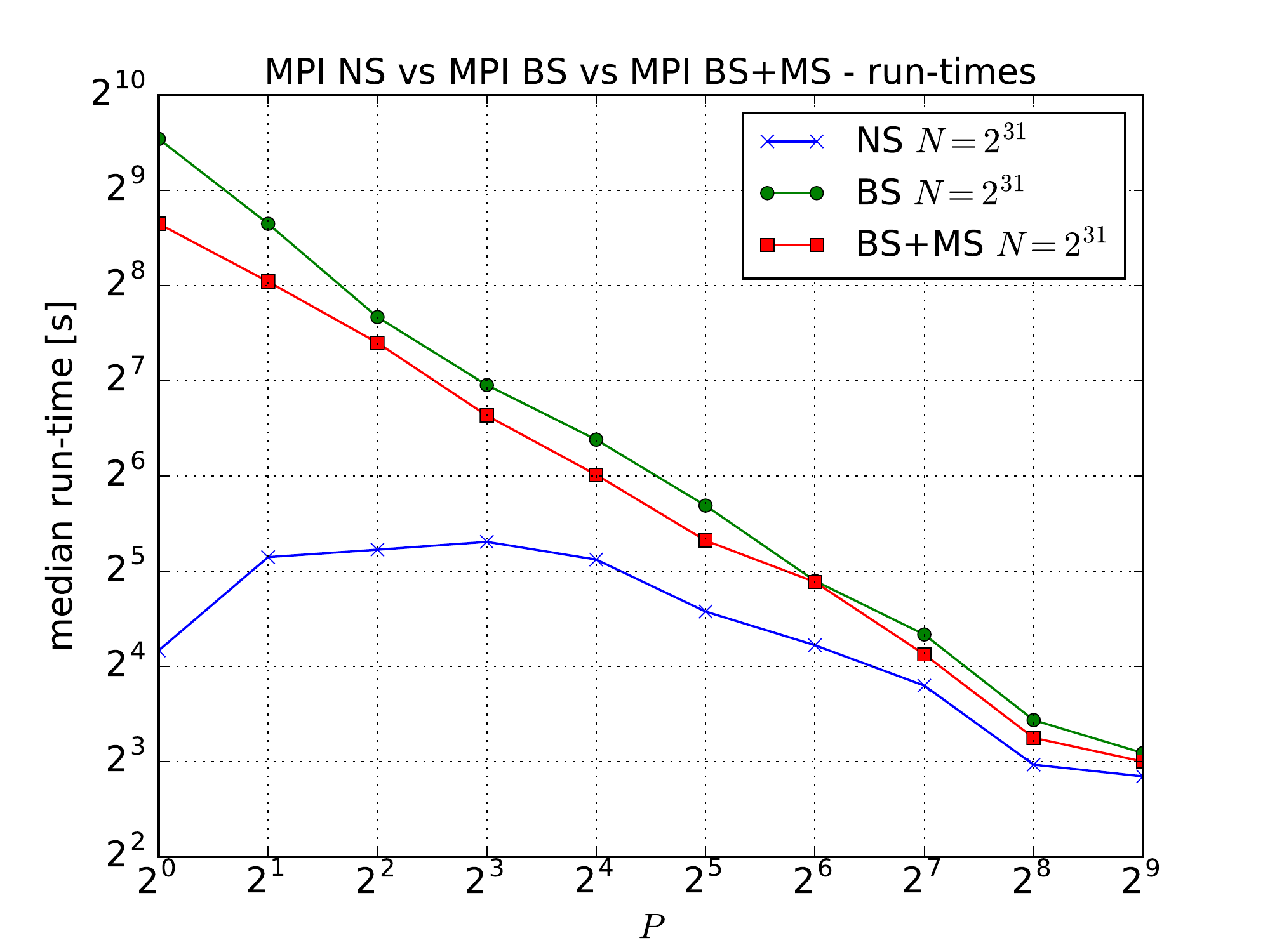}
					\caption[]%
					{{\small NS, BS and BS+MS for $N = 2^{31}$}}    
					\label{fig:ns_vs_bs_runtime_31}
				\end{subfigure}
				\hspace{1 cm}
				\begin{subfigure}[b]{0.33\textwidth}  
					\centering 
					\includegraphics[width=1\linewidth]{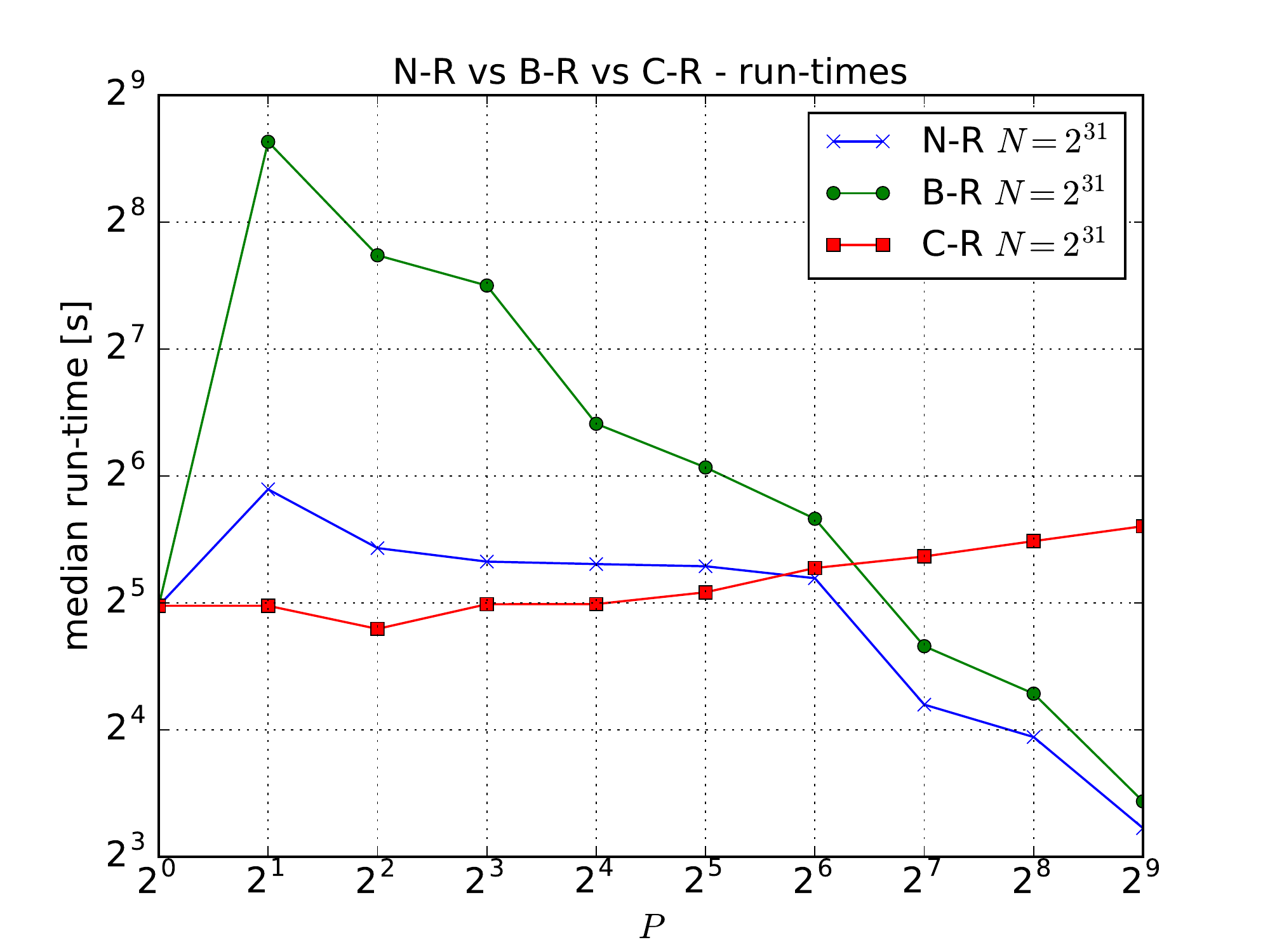}
					\caption[]%
					{{\small N-R, B-R and C-R for $N = 2^{31}$}}    
					\label{fig:redistribute_runtime_31}
				\end{subfigure}
	
				\caption[]
				{Bottleneck: median run-times vs $P$ for $N = 2^{10}, 2^{17}, 2^{24}, 2^{31}$} 
				\label{fig:bottleneck_results}
			\end{figure*}

			In this experiment, we use exactly the same strategy described in the previous section, since the required input for N-R, B-R and C-R is the same. The results are shown in Figure \ref{fig:bottleneck_results}. The results for redistribute with BS+MS are left out for brevity. 

			As we expected from theory, for $N \geq 2^{17}$ N-R is better than B-R overall and much faster for a small number of cores. However, the most important result is that N-R outperforms C-R at the theoretical minimum (which is $P = 32$) for some values of the dataset size $N$. This suggests that, as long as we use a parallel redistribute whose time complexity is equal to $O((log_2N)^2)$, we cannot outperform C-R for $P < 32$ nor have positive speed-up for the same values of $P$. In order to achieve this goal, a new algorithm with $O(log_2N)$ time complexity is needed. Sorting networks which achieve the theoretical lower bound have been proposed in \cite{Paterson1990}, \cite{Seiferas2009} which improve the original AKS sorting network presented in \cite{AKS}. These networks can also be rearranged to perform redistribute by substituting the comparators with balancers. However, they cannot be practically used because each atomic step requires a huge constant time $C$. The exact value of $C$ is unknown as it also depends on the network parameters but it seems to be in the order of thousands (e.g. $C = 6100$ in the best configuration in \cite{Paterson1990}). It can then be inferred that they cannot outperform $O((log_2N)^2)$ sorting networks such as Bitonic Sort for any practical $N$. In \cite{baddar2012designing} it has been estimated that a hypothetical $C = 87$ would require $N \geq 2^{173}$ to make AKS-like sorting networks faster than Bitonic Sort. Therefore, the infeasibility of this class of algorithms makes $O((log_2N)^2)$ redistribute on distributed memory systems a practical lower bound (although it cannot yet be considered a theoretical minimum).

			For $N = 2^{10}$, N-R does not scale and does not outperform C-R either. As we outlined in the previous section, for low values of $N$ the granularity is already too fine to observe any speed-up. 	

	\vspace{-5pt}
	\subsection{Worst-case speed-up} \label{subsec: Worst case speedup}
		In this section, we analyse three case studies, two for the Particle Filter and one example of an SMC Sampler. Depending on the chosen redistribute (N-R, B-R or C-R), we use the acronyms N-PF/B-PF/C-PF for the Particle Filter, and N-SMCS/B-SMCS/C-SMCS for the SMC Sampler. We consider a worst-case scenario which occurs when resampling is needed at every iteration and the time taken to perform Importance Sampling is small relative to the time taken to perform resampling. This section aims to achieve two goals. The first one is to demonstrate that the historic progress made in developing parallel implementations of Particle Filters can be translated to develop a parallel implementation of an SMC Sampler. The second one is to estimate the worst-case speed-up of our improved algorithm. 

		All experiments in this section are run for the same values of $N$ as were considered in the previous section. Each run-time is once again the median of $20$ runs, each one representing a simulation of $100$ iterations in the worst-case scenario and for increasing number of cores $P = 1, 2, 4, ..., 512$. 

		\subsubsection{Particle Filter on Econometrics} \label{subsubsec: PF_on_econ}

			In this experiment, we use Barkla and compare N-PF, B-PF and C-PF. We apply these algorithms to a stochastic volatility model which describes the evolution of the pound-to-dollar exchange rate between the 1st of October 1980 and the 28th of June 1985. This model has been used in \cite{Benchmarks_paper} to demonstrate the utility of advanced SMC methods, such as Block Sampling Particle Filters, over SIR Particle Filters.
			\begin{subequations}
				\begin{align}
					X_t = \phi X_{t-1} + \sigma V_t \label{eq: econ_model_state}\\
					Y_t = \beta \exp\left(\frac{X_t}{2}\right)W_t\label{eq: econ_model_measurement}
				\end{align}
			\end{subequations}

			(\ref{eq: econ_model_state}) and (\ref{eq: econ_model_measurement}) represent the model where the coefficients $\phi = 0.9731$, $\sigma = 0.1726$, $\beta = 0.6338$ (as selected in \cite{Benchmarks_paper}) and the noise terms for the state and the measurement are $V_t \sim \mathcal{N}(0,1)$ and $W_t \sim \mathcal{N}(0,1)$. The initial state is sampled as $X_0 \sim \mathcal{N}(0, \frac{\sigma^2}{1 - \phi^2})$. The particles are initially drawn from the prior distribution. The importance weights are simply equal to the likelihood $p\left(Y_t | X_t\right)$ since the dynamic model is used as the proposal.

		\subsubsection{Particle Filter on Bearing-Only Tracking} \label{subsubsec: PF_on_bearing_only_tracking}
			This experiment is focused on showing the performance of N-PF, B-PF and C-PF applied to a non-scalar model. The chosen example is the popular four-dimensional state model for a Bearing-Only Tracking problem, where the state is represented by the position and velocity of the tracked object. Both position and velocity are 2-dimensional physical quantities. This model was previously presented in several publications, such as in \cite{gordon}, and used in \cite{Benchmarks_paper} to test the Block Sampling SIR Filter. In accordance with \cite{Benchmarks_paper}, we consider the state to be composed of four elements denoted such that $\mathbf{X}_t = \left[X_t^0, X_t^1, X_t^2, X_t^3\right]$ where $X_t^0$, $X_t^2$ are position and $X_t^1$, $X_t^3$ are velocity. 

			The model is defined as follows:
			\begin{subequations}
				\begin{align}
					\mathbf{X}_t = \mathbf{A}\cdot\mathbf{X}_{t-1} + \mathbf{V}_t \\
					Y_t = \arctan \left(\frac{X_t^2}{X_t^0}\right) + W_t
				\end{align}
			\end{subequations}
			\noindent where the state transition matrix and the covariance are 
			\begin{align*}
				\mathbf{A} &= \begin{bmatrix}
					1 & \Delta & 0 & 0 \\
					0 & 1 & 0 & 0 \\
					0 & 0 & 1 & \Delta \\
					0 & 0 & 0 & 1
					\end{bmatrix}
				\hspace{0.1mm}
				&
				\mathbf{\Sigma} &= \begin{bmatrix}
					\frac{5\Delta^3}{3} & \frac{5\Delta^2}{2} & 0 & 0 \\
					\frac{5\Delta^2}{2} & 5\Delta & 0 & 0 \\
					0 & 0 & \frac{5\Delta^3}{3} & \frac{5\Delta^2}{2} \\
					0 & 0 & \frac{5\Delta^2}{2} & 5\Delta
				\end{bmatrix}
			\end{align*}
			The noise terms are $\mathbf{V}_t \sim \mathcal{N}\left(\mathbf{0}, \mathbf{\Sigma}\right)$ and $W_t \sim \mathcal{N}\left(0, 10^{-4}\right)$. The initial state $\mathbf{X}_0$ has the identity matrix as covariance and mean equal to the true initial simulated point of the system. The parameter $\Delta$ represents the sampling period and it is set to $\Delta = 1$ s.

		\subsubsection{Sampling using SMC samplers} \label{subsubsec: smcsamplers}
			We apply N-SMCS, B-SMCS and C-SMCS to sample from a static single-dimensional ($M = 1$) Student's t posterior distribution calculated as:
			\begin{equation} \label{eq: static multivariate student t}
				\begin{split}
					\pi(\mathbf{x}) =  \frac{\Gamma\left(\frac{\nu+1}{2}\right)}{\Gamma\left( \frac{\nu}{2}\right)\sqrt{(\nu\pi)}}
					\left(1 + \frac{1}{\nu} (\mathbf{x}-\mathbf{\mu})^2\right) ^{-\frac{\nu+1}{2}}
				\end{split}
			\end{equation} 
			\noindent where $\nu$ and $\mathbf{\mu}$ correspond to the degrees of freedom and the mean value respectively.

			In the experiment, the particles, as samples of $\mathbf{x}_t$, at the $t$-th iteration are generated using random walk as the proposal distribution, $\mathbf{x}_t \sim \mathcal{N}(\mathbf{x}_{t-1}, \epsilon)$. The backward kernel is naively selected to emulate MCMC such that $L_t(\mathbf{x}_{t-1}|\mathbf{x}_t) =  q_t(\mathbf{x}_t|\mathbf{x}_{t-1})$. We then anticipate that a better choice of $L_t(\mathbf{x}_{t-1}|\mathbf{x}_t)$ can positively impact our estimate. The recycling described in Section \ref{subsubsec: smcs} is also used. 

		\subsubsection{Worst-case results} \label{subsubsec: worst case results}
\begin{figure*}[!htp]
				\centering
				\begin{subfigure}[b]{0.32\textwidth}
					\centering
					\includegraphics[width=1\linewidth]{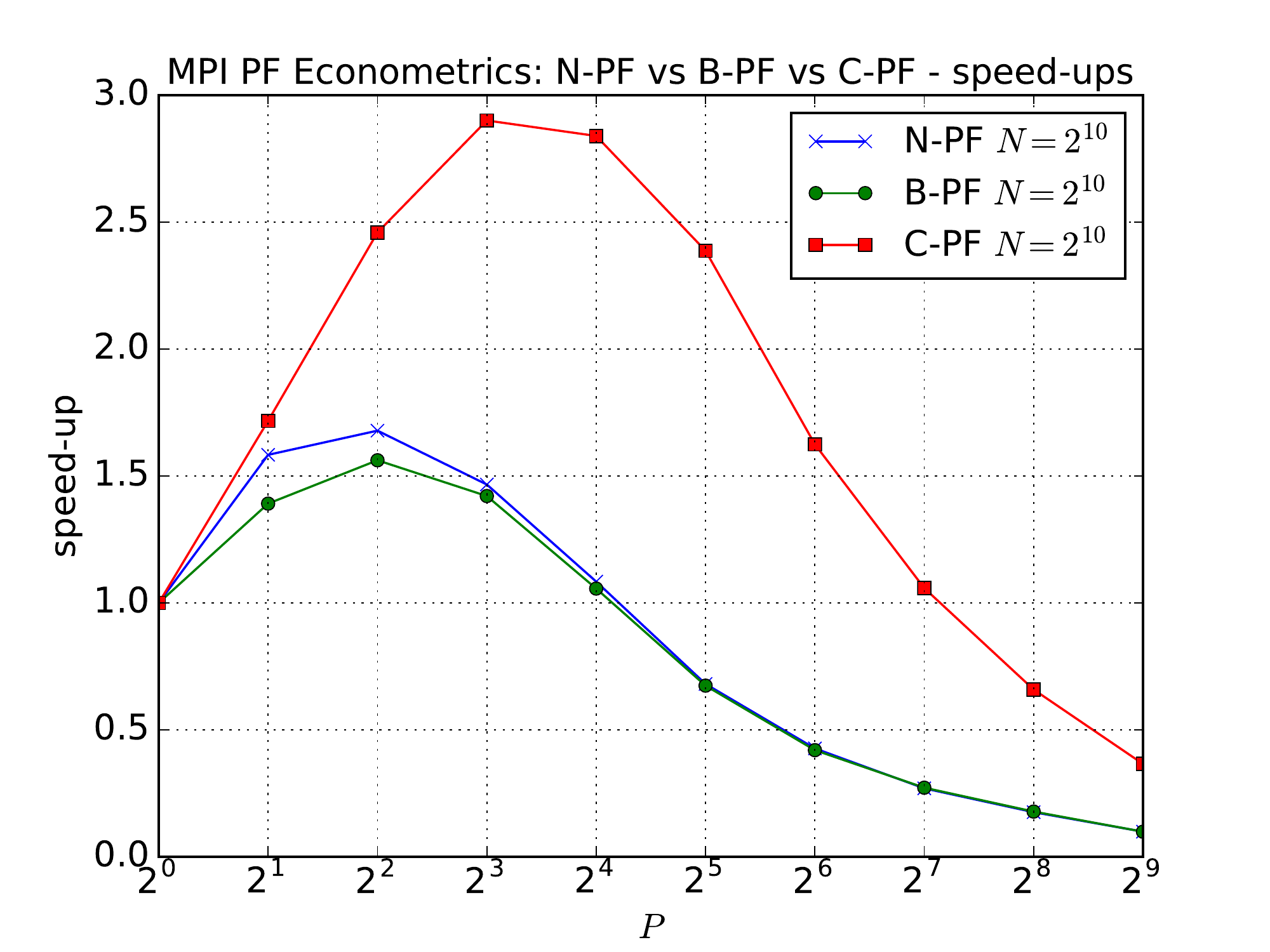}
					\caption[]%
					{{\small Econometrics for $N = 2^{10}$}}    
					\label{fig:econ_runtime_10}
				\end{subfigure}
				\hfill
				\begin{subfigure}[b]{0.32\textwidth}  
					\centering 
					\includegraphics[width=1\linewidth]{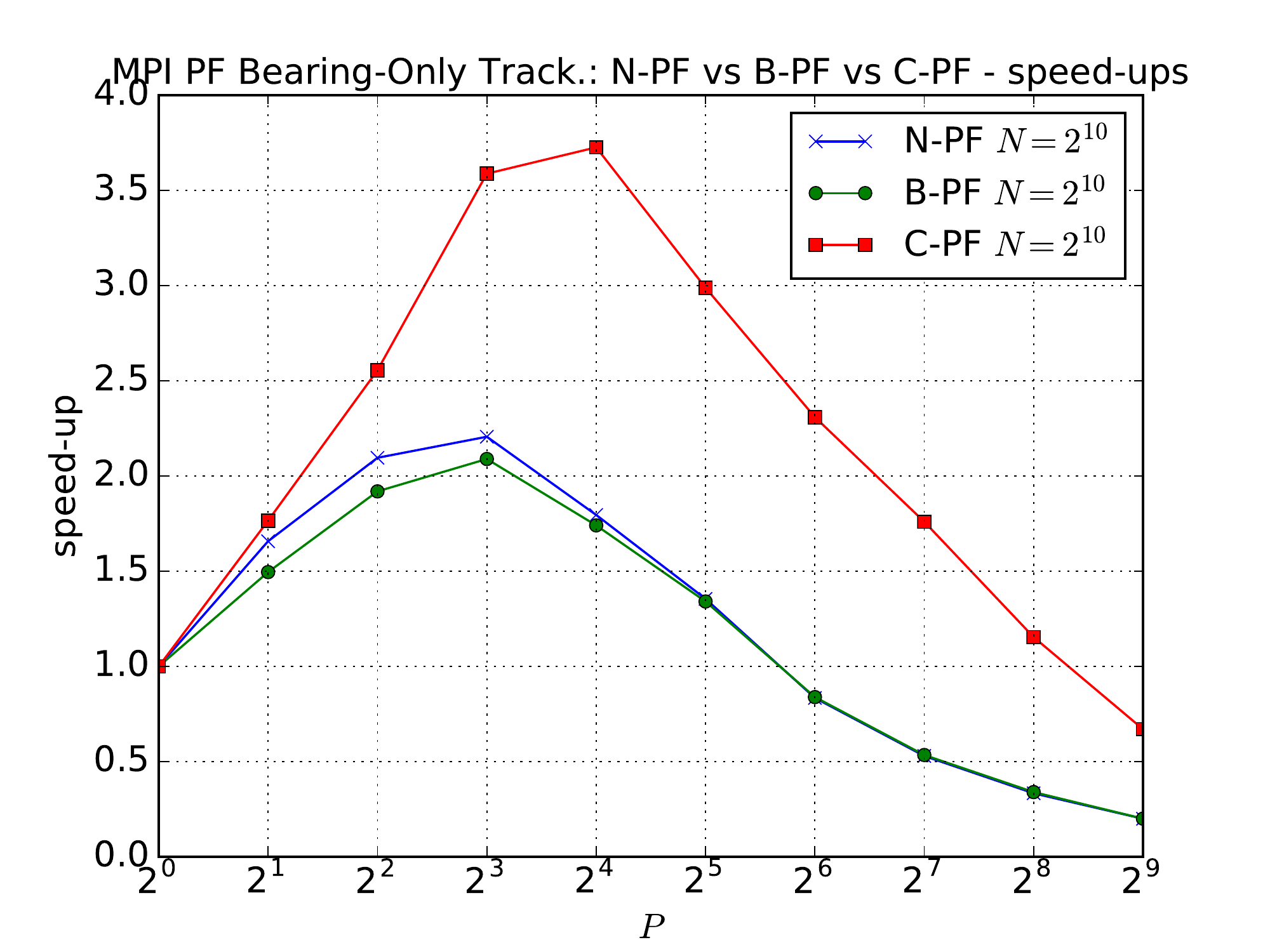}
					\caption[]%
					{{\small Bearing-Only Track. for $N = 2^{10}$}}    
					\label{fig:bo_runtime_10}
				\end{subfigure}
				\hfill
				\begin{subfigure}[b]{0.32\textwidth}   
					\centering 
					\includegraphics[width=1\linewidth]{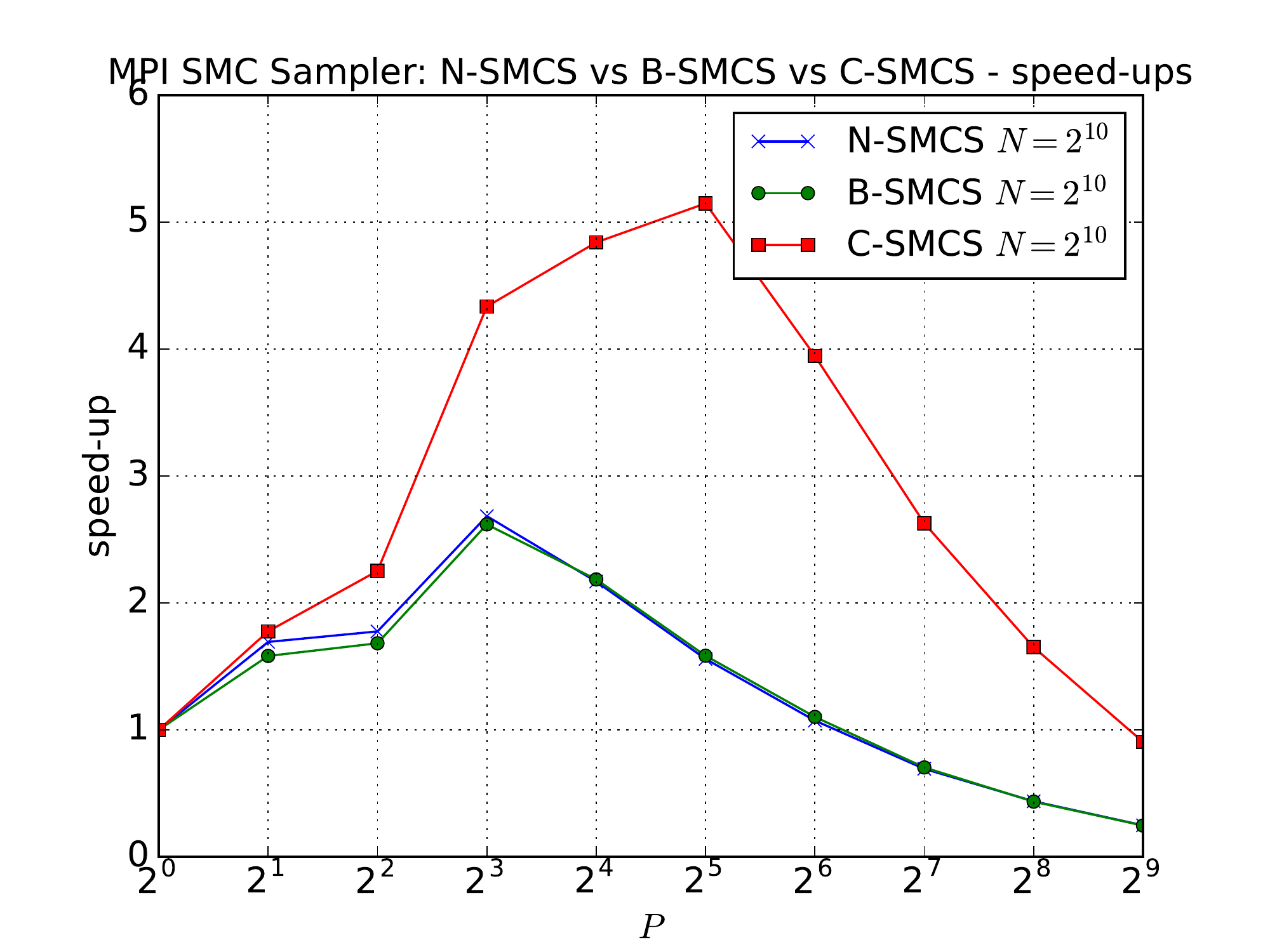}
					\caption[]%
					{{\small Synthetic SMC Sampler for $N = 2^{10}$}}    
					\label{fig:smcs_runtime_10}
				\end{subfigure}
				\vskip\baselineskip
	
				\centering
				\begin{subfigure}[b]{0.32\textwidth}
					\centering
					\includegraphics[width=1\linewidth]{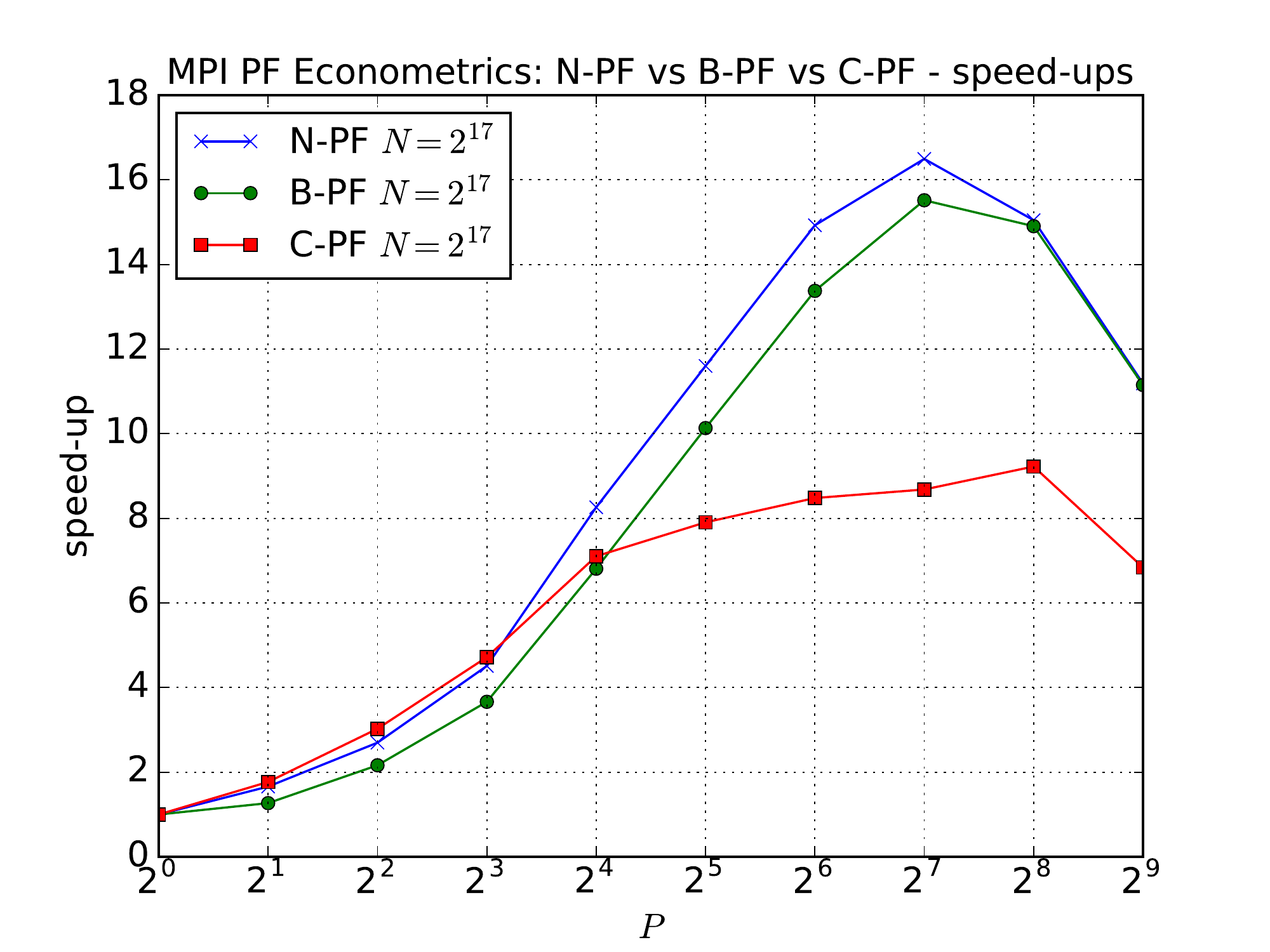}
					\caption[]%
					{{\small Econometrics for $N = 2^{17}$}}    
					\label{fig:econ_runtime_17}
				\end{subfigure}
				\hfill
				\begin{subfigure}[b]{0.32\textwidth}  
					\centering 
					\includegraphics[width=1\linewidth]{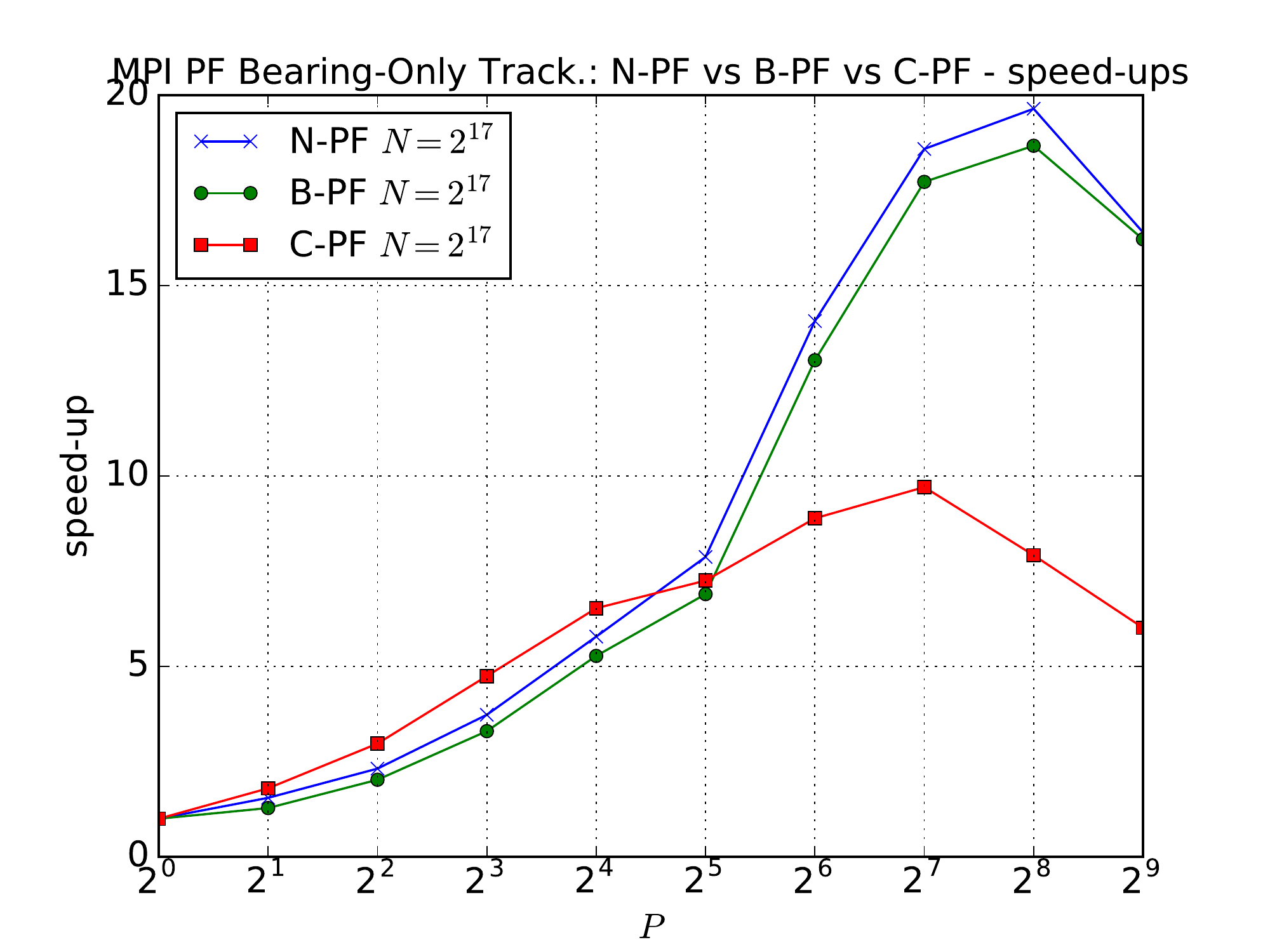}
					\caption[]%
					{{\small Bearing-Only Track. for $N = 2^{17}$}}    
					\label{fig:bo_runtime_17}
				\end{subfigure}
				\hfill
				\begin{subfigure}[b]{0.32\textwidth}   
					\centering 
					\includegraphics[width=1\linewidth]{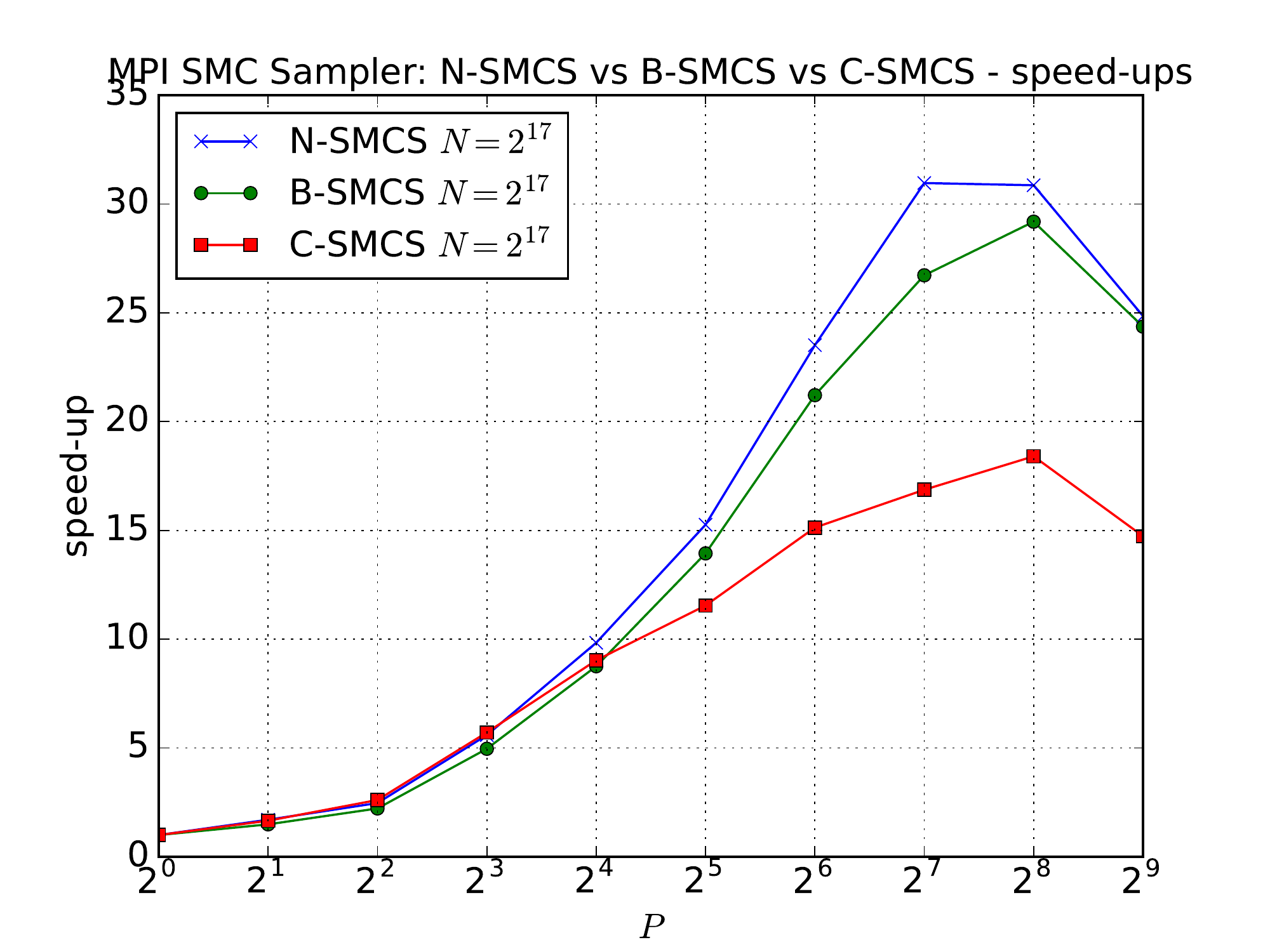}
					\caption[]%
					{{\small Synthetic SMC Sampler for $N = 2^{17}$}}    
					\label{fig:smcs_runtime_17}
				\end{subfigure}
				\vskip\baselineskip
	
				\centering
				\begin{subfigure}[b]{0.32\textwidth}
					\centering
					\includegraphics[width=1\linewidth]{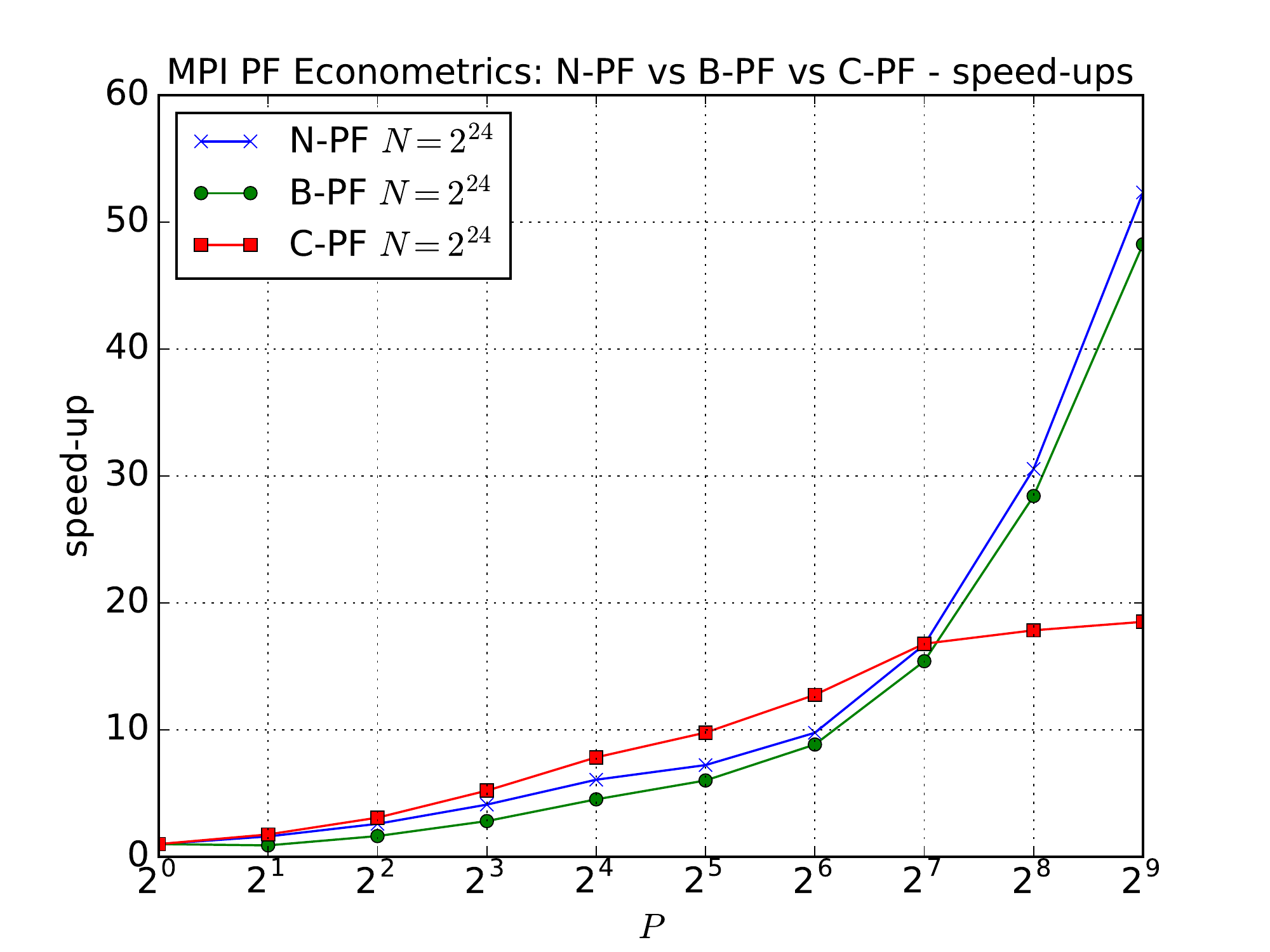}
					\caption[]%
					{{\small Econometrics for $N = 2^{24}$}}    
					\label{fig:econ_runtime_24}
				\end{subfigure}
				\hfill
				\begin{subfigure}[b]{0.32\textwidth}  
					\centering 
					\includegraphics[width=1\linewidth]{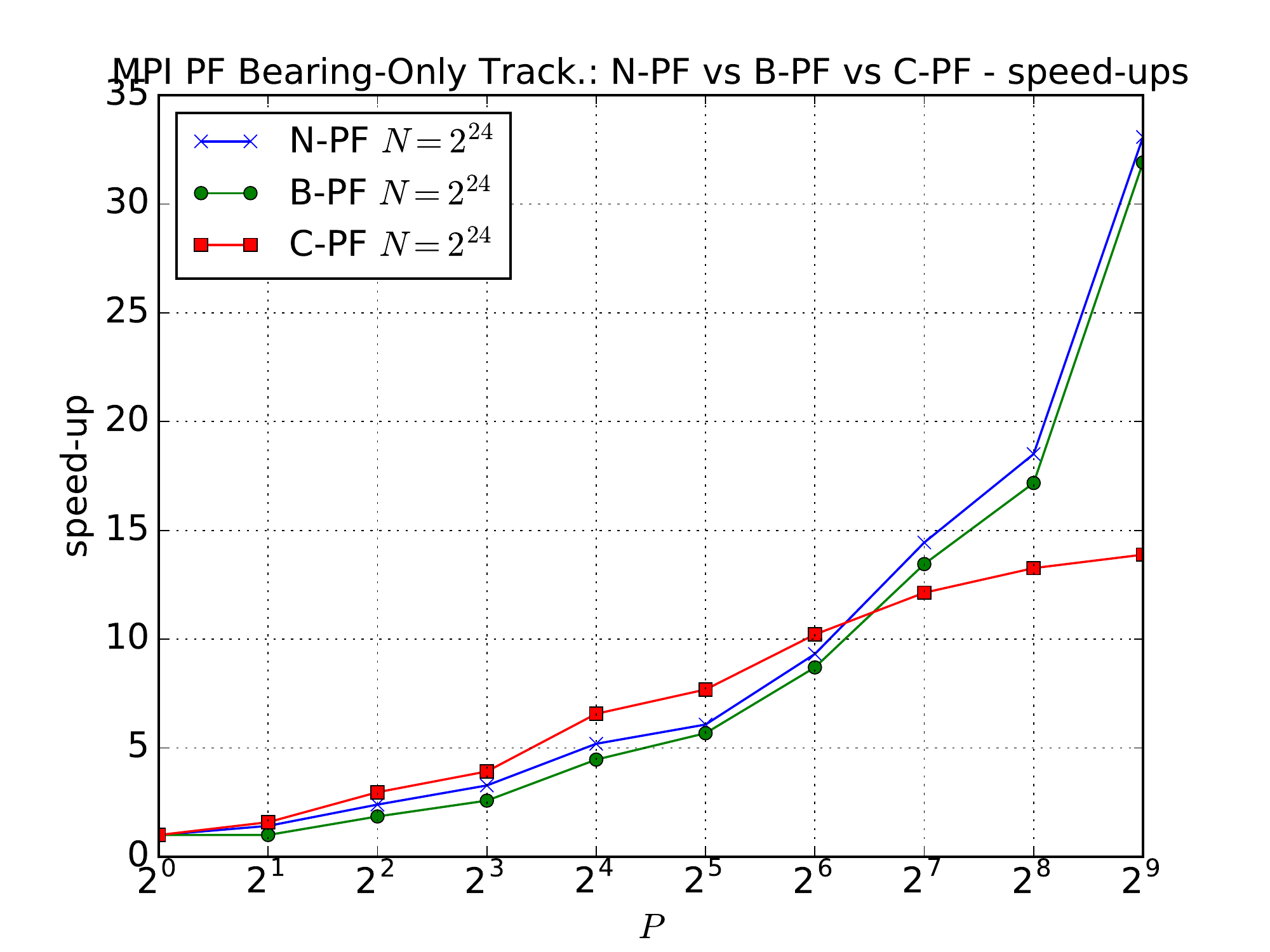}
					\caption[]%
					{{\small Bearing-Only Track. for $N = 2^{24}$}}    
					\label{fig:bo_runtime_24}
				\end{subfigure}
				\hfill
				\begin{subfigure}[b]{0.32\textwidth}   
					\centering 
					\includegraphics[width=1\linewidth]{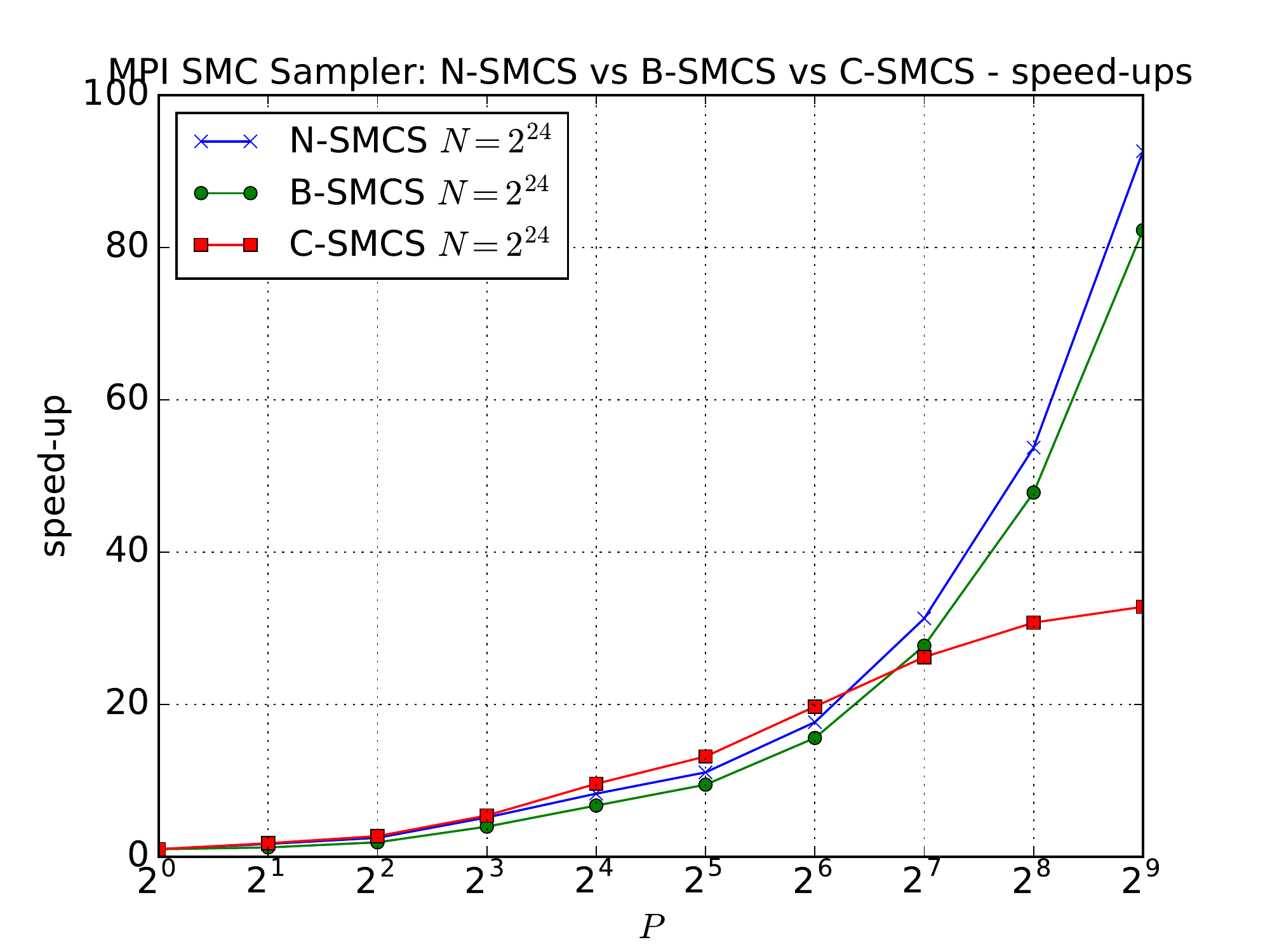}
					\caption[]%
					{{\small Synthetic SMC Sampler for $N = 2^{24}$}}    
					\label{fig:smcs_runtime_24}
				\end{subfigure}
				\vskip\baselineskip
	
				\centering
				\begin{subfigure}[b]{0.32\textwidth}
					\centering
					\includegraphics[width=1\linewidth]{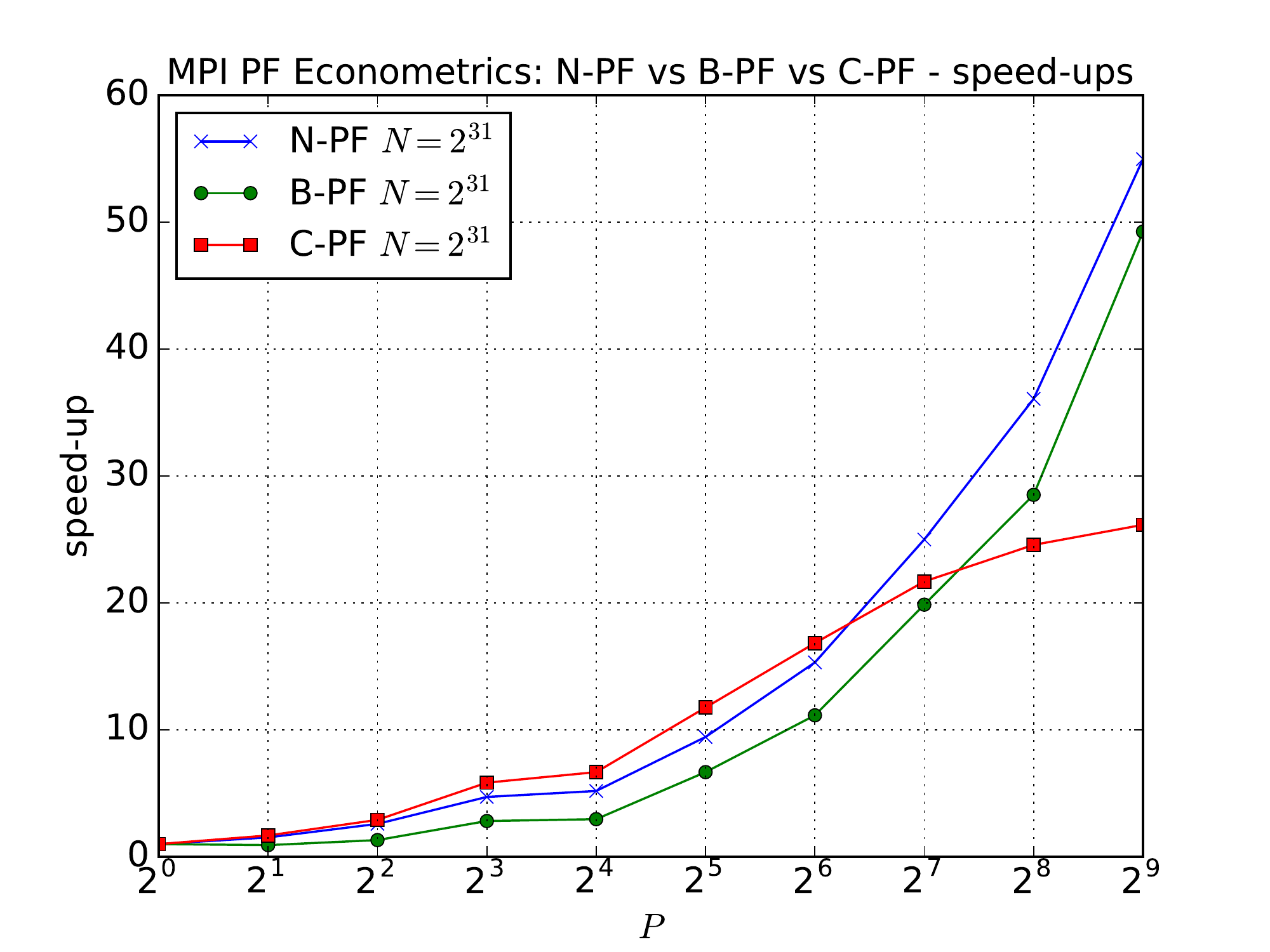}
					\caption[]%
					{{\small Econometrics for $N = 2^{31}$}}    
					\label{fig:econ_runtime_31}
				\end{subfigure}
				\hfill
				\begin{subfigure}[b]{0.32\textwidth}  
					\centering 
					\includegraphics[width=1\linewidth]{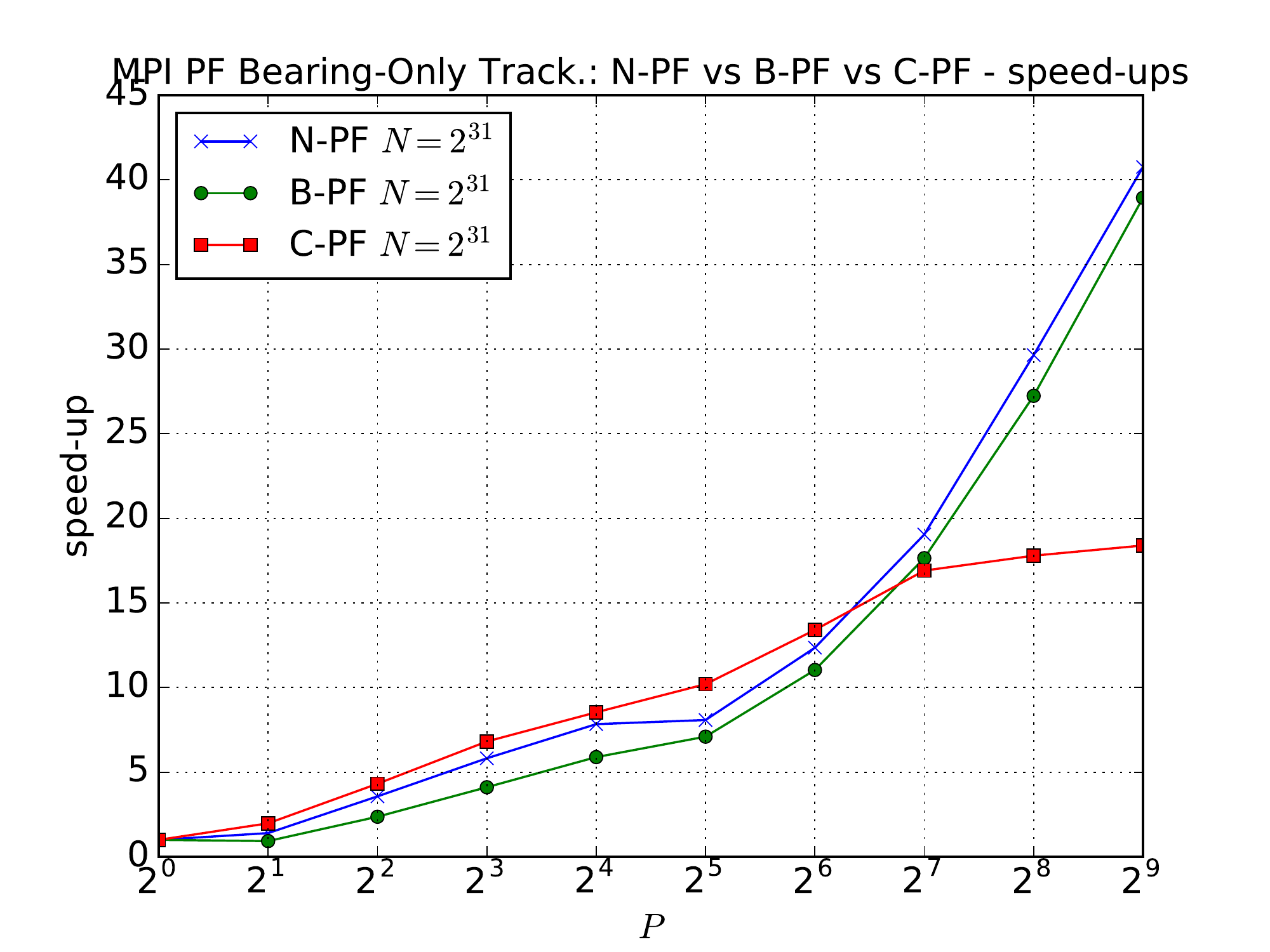}
					\caption[]%
					{{\small Bearing-Only Track. for $N = 2^{31}$}}    
					\label{fig:bo_runtime_31}
				\end{subfigure}
				\hfill
				\begin{subfigure}[b]{0.32\textwidth}   
					\centering 
					\includegraphics[width=1\linewidth]{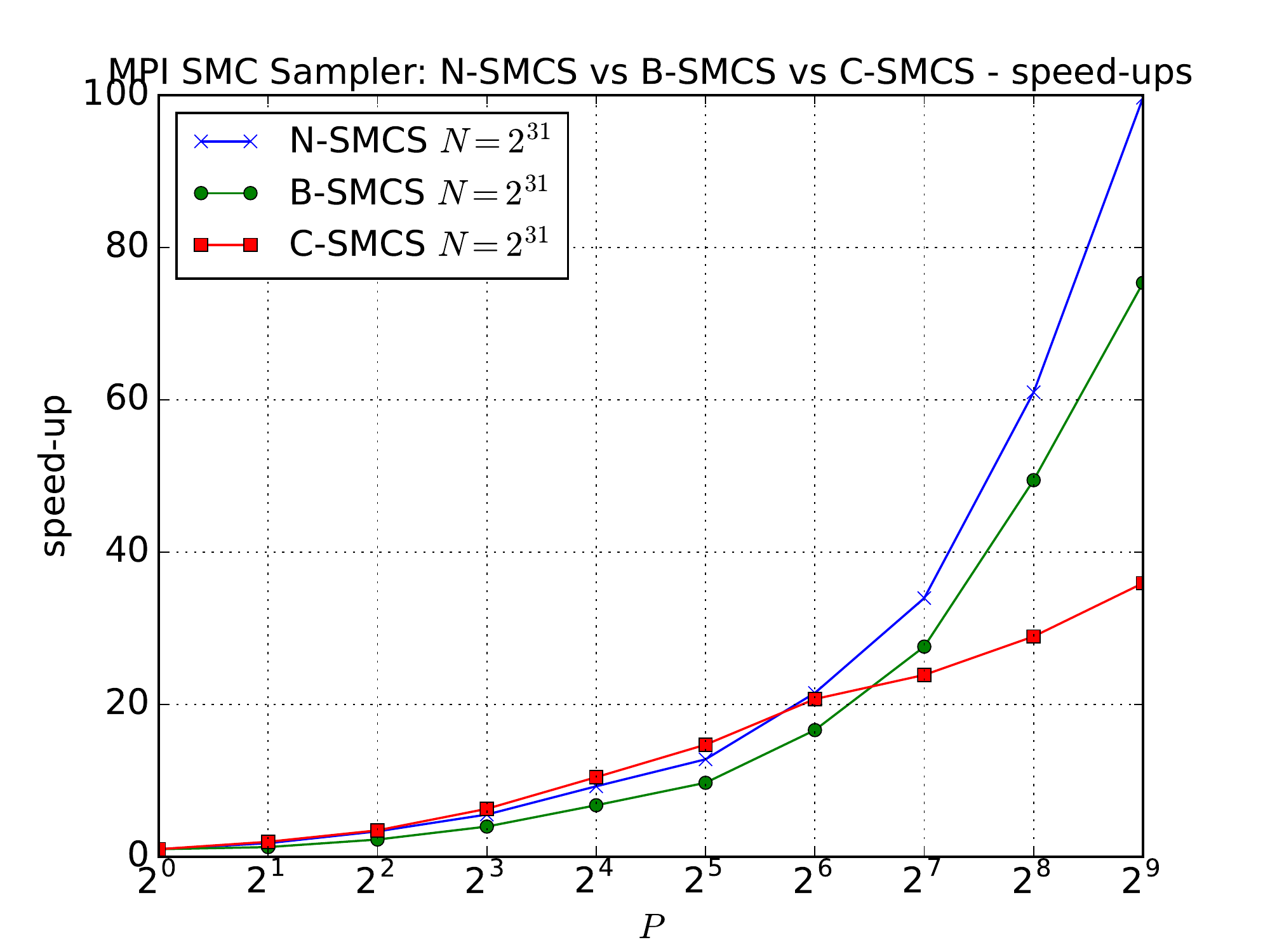}
					\caption[]%
					{{\small Synthetic SMC Sampler for $N = 2^{31}$}}    
					\label{fig:smcs_runtime_31}
				\end{subfigure}
	
				\caption[]
				{SMC methods: speed-ups vs $P$ for $N = 2^{10}, 2^{17}, 2^{24}, 2^{31}$} 
				\label{fig:SMC_methods_results}
			\end{figure*}

			In this section we provide the results of the experiments described in Sections \ref{subsubsec: PF_on_econ}, \ref{subsubsec: PF_on_bearing_only_tracking} and \ref{subsubsec: smcsamplers} which are shown in Figure \ref{fig:SMC_methods_results} for $N = 2^{10}, 2^{17}, 2^{24}, 2^{31}$ respectively. Since that N-R, B-R and C-R have the same baseline, we show the speed-ups instead of the run-times as, in this case, these speed-ups can also prove which algorithm is faster.

			As we can see, for $N = 2^{10}$ N-PF/N-SMCS does not outperform C-PF/C-SMCS. The reasons are the granularity and the theoretical time complexity of N-R as we explained in Section \ref{subsec: bottleneck}. In contrast, C-PF/C-SMCS can keep scaling for a relatively low $P$, until redistribute emerges as the bottleneck. However, since modern applications need a large number of particles ($2^{10}$ is just $1024$), we are not discouraged by these limitations. 

			For $N \geq 2^{17}$, computation becomes dominant over communication and both N-PF/N-SMCS and B-PF/B-SMCS can scale for much larger values of $P$ and can both outperform C-PF/C-SMCS. For this range of $N$, using Nearly Sort instead of Bitonic Sort makes the SMC method faster for any number of cores and up to twice as fast as using Bitonic Sort for low values of $P$. The gap between these two approaches also increases with $N$. N-PF/N-SMCS runs faster than the solution with C-R in the range $32 \leq P \leq 128$ and it can be as much as approximately $3$ times faster for $P = 512$ cores. 

			Overall, we can say that for a fixed $N$, changing the model in the Particle Filter (i.e. whether we consider stochastic volatility or Bearing-Only Tracking) or switching from Particle Filters to SMC Samplers gives roughly the same trend. This proves that the improvements that have been demonstrated in the context of Particle Filters can directly be translated to the context of SMC Samplers.

			As we can see, the minimum worst-case speed-up is about $40$ and occurs in the context of the Bearing-Only Tracking. On the other hand, the maximum worst-case speed-up can be up to $100$: this occurs in the context of the SMC Sampler. The efficiency of N-SMCS with respect to the maximum speed-up is indeed significantly higher than for the Particle Filter. This is due to the different way particles and weights are calculated in the SMC Sampler. For example, (\ref{eq: static multivariate student t}) is more computationally intensive than (\ref{eq: econ_model_state}), and the likelihood calculation in the SMC Sampler is more computationally demanding than the likelihood in the Particle Filters (this is because of the need to compute the $L$-kernel). Therefore, resampling accounts for a lower percentage of the entire workload in the SMC Sampler than it does in the Particle Filter. The resampling step is no longer such a significant bottleneck for a low number of cores. This is discussed in more detail in the next section. 

	\vspace{-5pt}
	\subsection{Maximum speed-up} \label{subsec: PF_on_multi_sensor_bo_tracking}
		All the previous experiments use relatively simple proposal distributions and likelihoods. However, in real problems, these two tasks are likely to be much more complicated (e.g. they may involve non-linear systems or Partial Differential Equations etc.). In the next experiment, we investigate the impact that a more computationally intensive Importance Sampling step has on the maximum speed-up. 

		In order to simulate this scenario, we adjust the experiment described in Section \ref{subsubsec: PF_on_bearing_only_tracking} by using $D > 1$ sensors spread over the Cartesian plane. This practice is also common in real applications to make the estimate more accurate (since the triangulation observability criterion is satisfied \cite{multi_sensors_bo_track}). The measurement is now a $D$-dimensional measurement vector:
		\begin{equation}
			\mathbf{Y}_t = \arctan \left(\frac{X_t^2 - \tilde{y_k}}{X_t^0 - \tilde{x_k}}\right) + \mathbf{W}_t \ \ \forall k = 1, ..., D
		\end{equation}
		where $(\tilde{x_k}, \tilde{y_k})$ is the position of the $k$-th sensor with respect to the target. The state equation remains the same as is described in Section \ref{subsubsec: PF_on_bearing_only_tracking} such that $M$ is unchanged. We consider $N = 2^{24}$. The maximum speed-up efficiency for each $D$ is estimated as the speed-up for $P = 512$ vs the ideal speed-up for the same $P$. We increase $D$ until we have at least $50\%$ efficiency. For each $D$ we also report the percentage of the total workload that Importance Sampling accounts for when the run-time of redistribute is at its peak, i.e. for $P = 2$. 

		As we can observe in Figure \ref{fig:MPI_multi_sensors}, a more computationally intensive Importance Sampling step leads to higher speed-ups. The speed-up for $D = 360$ is indeed about $7.3$ times the speed-up for $D = 1$ (which corresponds to the experiment in Section~\ref{subsubsec: PF_on_bearing_only_tracking}). Therefore, in these problems, the bottleneck for a low number of cores is likely to be the Importance Sampling step and not resampling. However, when $P = N$, Importance Sampling has $O(1)$ time complexity while resampling has complexity of  $O((log_2N)^2)$. In other words, since resampling always emerges as the bottleneck for a sufficiently high level of parallelism, it is crucial to use a parallelisable redistribute such that we can achieve near-linear speed-ups for higher $P$.

	\vspace{-5pt}	
	\subsection{Space Complexity} \label{subsec: unfeasibility}
		N-R and B-R, have both scalable space complexity equal to $O(M \cdot \frac{N}{P})$. However, C-R has constant space complexity equal to $O(M \cdot N)$ \cite{Alessandro}: one core is in charge of collecting the particles, performing the routine locally and then distributing the new population back to the other cores.

		The main side effect is that when the available memory in each node is insufficient to store all the necessary data for $P = 1$, we cannot run C-R for any $P$. In contrast, even for very large values of $N$, we can always run N-R or B-R as long as each node has enough memory for its data. In order to show this problem, we repeat the experiment described in Section \ref{subsubsec: PF_on_bearing_only_tracking} on Chadwick (which has a $64$ GB memory in each node, i.e. less than Barkla's $384$ GB per node). Figure \ref{fig:MPI_PF_bearing_runtime_chadwick} shows the measured run-times for $N = 2^{31}$ (speed-ups are not provided since the baseline is impossible to run due to space complexity limitations). The results for $N \in \{2^{10}, 2^{17}, 2^{24}\}$ are left out for brevity since they resemble the results in Figure \ref{fig:SMC_methods_results}. The total absence of a curve for C-PF or the missing points for N-PF occur because of an mpirun abort (which happens when we request more memory than the node has). As we can see we need at least $P = 64$ cores to run N-PF while it is never possible with C-PF with $N = 2^{31}$. 

		We can conclude that N-PF/N-SMCS outperforms B-PF/B-SMCS for any $P$ and outperforms C-PF/C-SMCS for $P \geq 32$ cores. Furthermore, it is always possible to run N-PF/N-SMCS while C-PF/C-SMCS may be impossible to run for high $N$.

	\vspace{-5pt}
	\subsection{SMC Sampler vs Metropolis-Hastings}
		\label{subsec: smcs_vs_mcmc}
		\subsubsection{Description} \label{subsubsec: smcs_vs_mcmc_description}	
			As we have anticipated in Section \ref{subsec: smsc_vs_mh}, this experiment aims to achieve two goals. We first want to prove that a $P$-core implementation of the SMC Sampler can achieve a lower run-time than a single-chain Metropolis-Hastings when both algorithms draw the same number of samples in total (see below). Then we want to prove that the extra speed-up that $P$ cores provide can make an SMC Sampler more accurate than Metropolis-Hastings, since as $P$ increases an SMC Sampler can perform more iterations over the same time span.

			The first part of the experiment is done by comparing the run-time of both algorithms for the same workload such that:
			\begin{equation} \label{eq: smcs_vs_mcmc_first}
				T_{MH} = N \times T_{SMC}
			\end{equation}

			To investigate the second issue, we primarily need to know the inter-task speed-up $SU_P$ which $P$ cores can provide, keeping $N$ fixed. We estimate $SU_P$ from the first part of the experiment. Then we compare the algorithms in terms of accuracy over the same computational time which happens when:
			\begin{equation} \label{eq: smcs_vs_mcmc_second}
				T_{MH} = N \times T_{SMC} \times SU_P, \thickspace P = 1, 2, 4, ...
			\end{equation}
			In other words, the SMC Sampler will run for $SU_P$-times more iterations (or less if $0 \leq SU_P \leq 1$) over the same run-time. 

			The number of samples, $N$, and the number of cores, $P$, are the same as in the previous experiments and we will use $T_{SMC} \in \{100, 1000, 10000\}$. $T_{MH}$ is constant, independent of $P$ and always picked using (\ref{eq: smcs_vs_mcmc_second}) for $P = 1$. The SMC Sampler is once again assessed in the worst-case setting when resampling is needed at every iteration.

		\subsubsection{Results} \label{subsubsec: smcs_vs_mcmc_results}

			\begin{figure*}
				\captionsetup{justification=centering}
				\centering
				\minipage[b]{0.33\textwidth}
					\includegraphics[width=\textwidth]{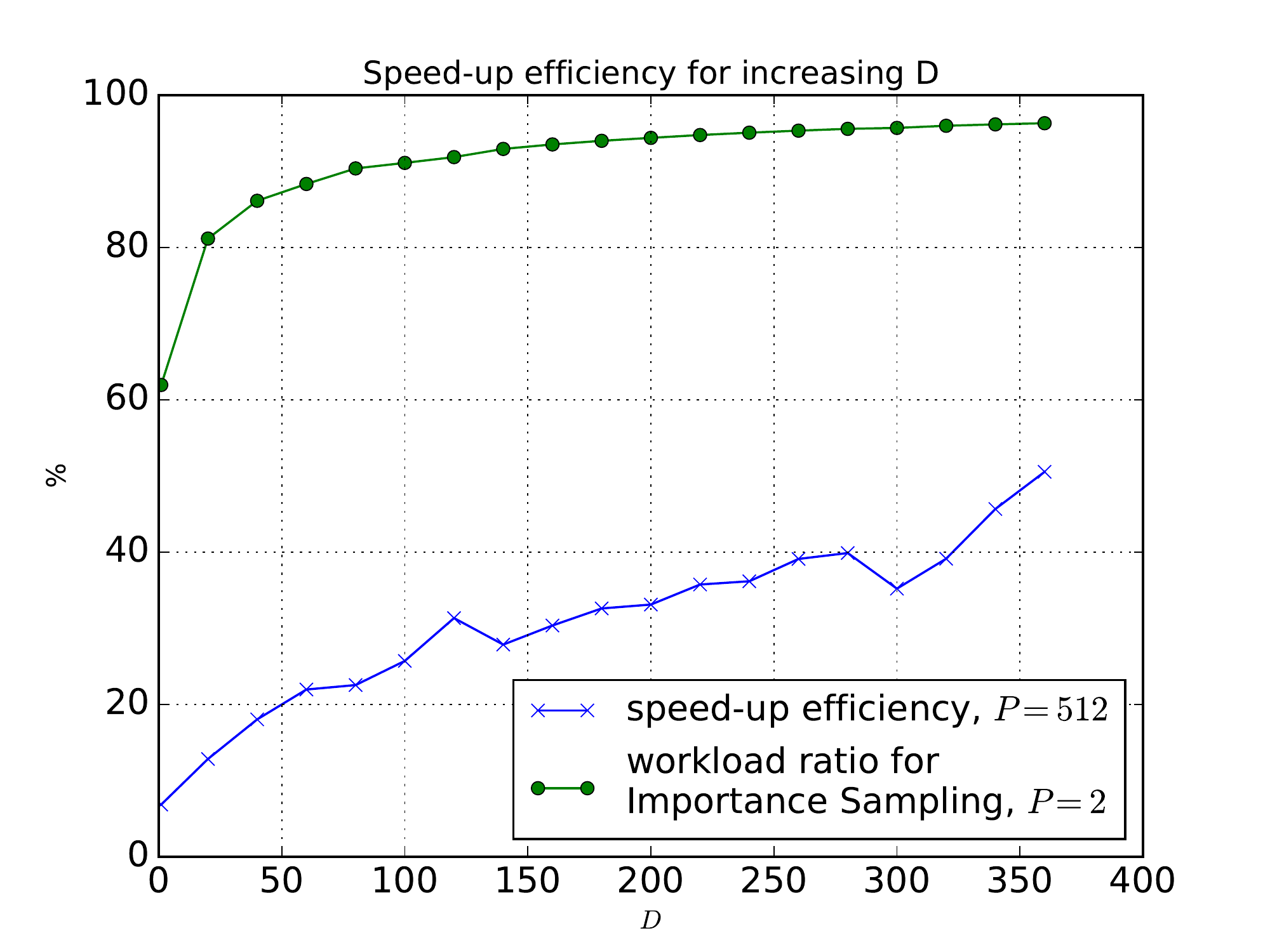}
					\caption{Multi Sensors: max \\speed-up efficiency for increasing $D$}   
					\label{fig:MPI_multi_sensors}
				\endminipage
				\centering
				\minipage[b]{0.33\textwidth}
					\includegraphics[width=\textwidth]{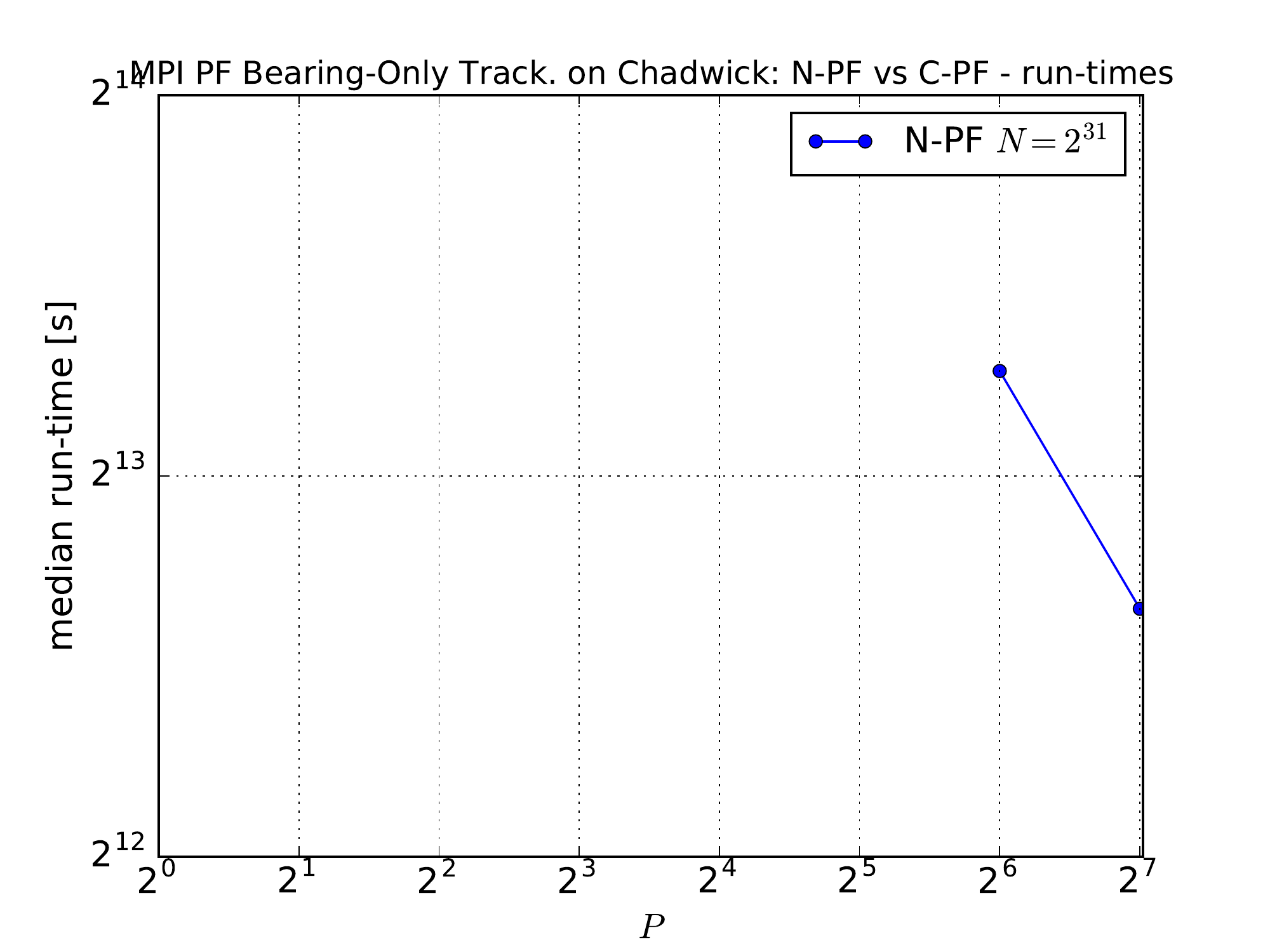}
					\caption{Bearing-Only Track.\\ for $N = 2^{31}$ on Chadwick}   
					\label{fig:MPI_PF_bearing_runtime_chadwick}
				\endminipage
				\centering
				\minipage[b]{0.33\textwidth}
					\includegraphics[width=\textwidth]{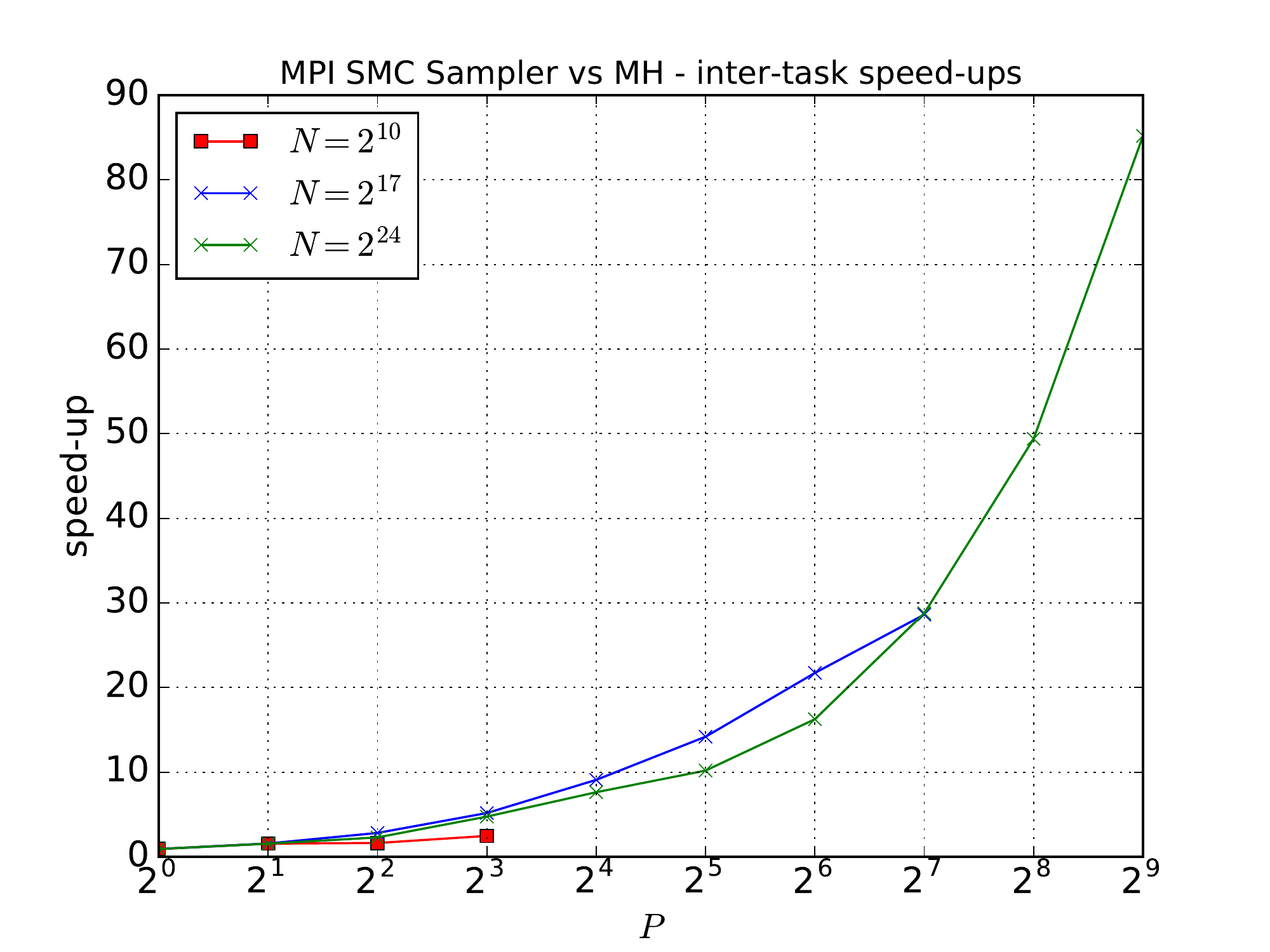}
					\caption{N-SMCS vs MH: inter-task speed-ups vs $P$ for $T_{SMC} = 100$}
					\label{fig:MPI_smcs_vs_mcmc_speedups}
				\endminipage
			\end{figure*}

			Figure \ref{fig:MPI_smcs_vs_mcmc_speedups} shows the inter-task speed-up between SMC Sampler and Metropolis-Hastings for the same workload (see (\ref{eq: smcs_vs_mcmc_first})), after having set $T_{SMC} = 100$. We calculate that a single-core implementation of the SMC Sampler is slightly slower (typically by $8 \%$) than Metropolis-Hastings. This means that an SMC Sampler running on a cluster of nodes could be much faster than Metropolis-Hastings. As we can see, for high values of $N$ (which as we have seen lead to larger speed-ups) the SMC Sampler can be up to $85$ times faster than Metropolis-Hastings. 

			\begin{figure*}[!htp]
				\centering
				\minipage{0.33\textwidth}
					\centering
					\includegraphics[width=1\linewidth]{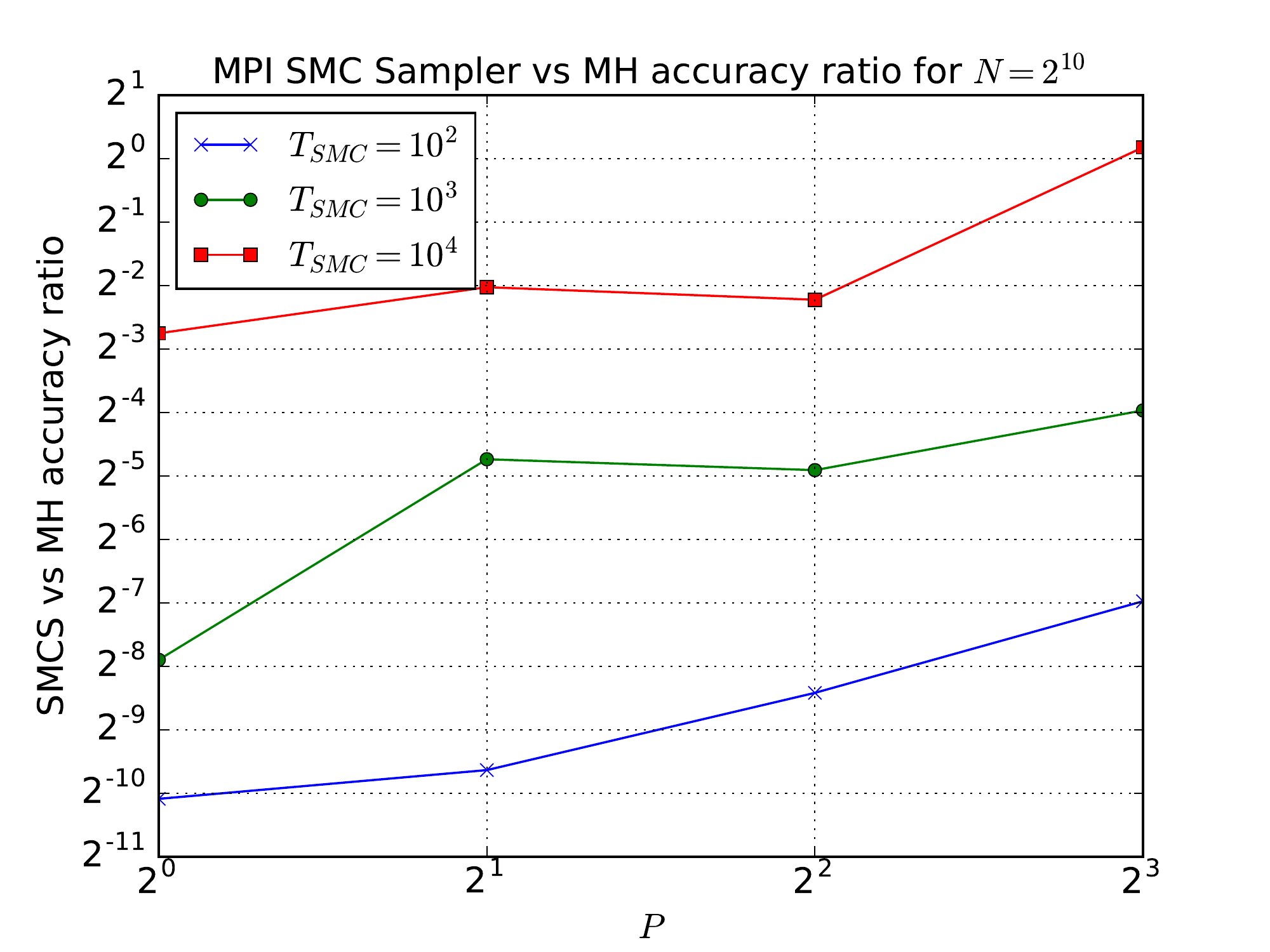}
					\\ (a) $N = 2^{10}$
				\endminipage  
				\minipage{0.33\textwidth}
					\centering 
					\includegraphics[width=1\linewidth]{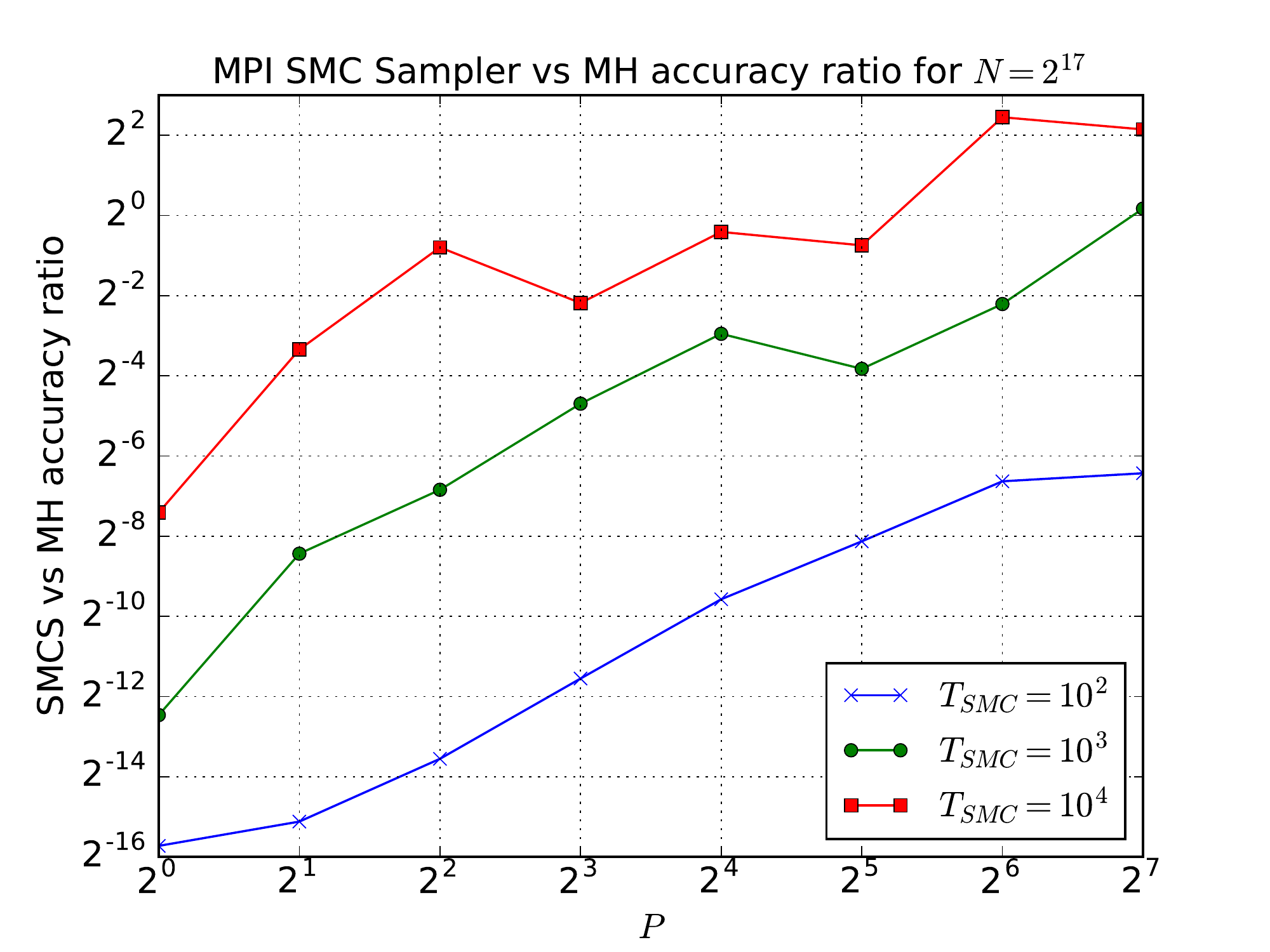}
					\\ (b) $N = 2^{17}$
				\endminipage 
				\minipage{0.33\textwidth}
					\centering
					\includegraphics[width=1\linewidth]{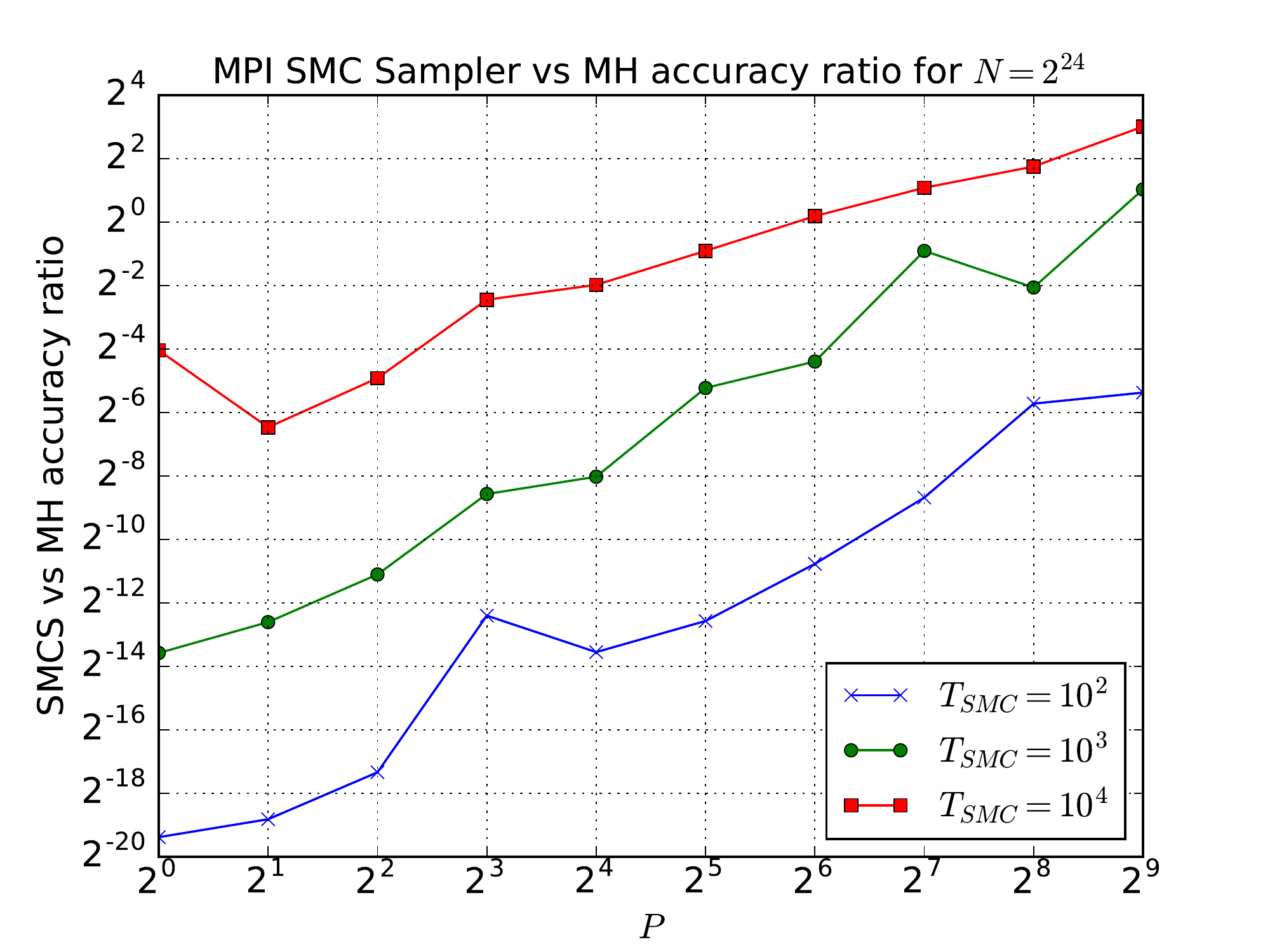}
					\\ (c) $N = 2^{24}$
				\endminipage 
					\caption{SMC Sampler vs Metropolis-Hastings accuracy ratios}
					\label{fig: SMCS_vs_MCMC}
			\end{figure*}

			In the second part of the experiment we compare the accuracy (expressed as Root Mean Squared Error (RMSE)) for high speed-ups when both algorithms run for the same time span (see (\ref{eq: smcs_vs_mcmc_second})). To make the run-time equivalent we set $SU_P$ to the values shown in Figure \ref{fig:MPI_smcs_vs_mcmc_speedups}. Figure \ref{fig: SMCS_vs_MCMC} shows the accuracy ratios between the algorithms vs $P$ for increasing $N$ or $T_{SMC}$. The baseline (and numerator of the accuracy ratio) is the accuracy of Metropolis-Hastings (see Table \ref{tab: mh_rsme}). As we can see, for low values of $T_{SMC}$, the SMC Sampler does not outperform Metropolis-Hastings as the gap for $P = 1$ is initially too big. However, while an SMC Sampler is less accurate, the relative benefit of using Metropolis-Hastings reduces when $P$ increases. Combinations of bigger values for $N$ or $T_{SMC}$ lead to comparable gains in accuracy when $SU_P$ is maximised (see pairs: $N = 2^{10}$, $T_{SMC} = 10^4$; $N = 2^{17}$, $T_{SMC} = 10^3$; $N = 2^{24}$, $T_{SMC} = 10^3$). In the end, for even higher values of $T_{SMC}$ or $N$ the initial gap at the baseline is lower such that, when $SU_P$ increases sufficiently, the SMC Sampler finally outperforms Metropolis-Hastings. Figures \ref{fig: SMCS_vs_MCMC}c proves that the RMSE of the SMC Sampler can be up to approximately $8$ times lower. If we consider that the standard deviation $\sigma$ for Metropolis-Hastings scales as $O(1/\sqrt{N})$, we can infer that the ideal improvement in accuracy for $P = 512$ cores would be:
			\begin{equation}
				\frac{\sigma_{P=512}}{\sigma_{P=1}} = \frac{\sqrt{N}}{\sqrt{PN}} = \frac{1}{\sqrt{512}} \approx \frac{1}{22.63}
			\end{equation}
			This would occur only if we could trivially parallelise a single-chain Metropolis-Hastings and observe linear speed-up, which means that the proposed MPI SMC Sampler already achieves approximately $35\%$ efficiency with respect to the ideal scenario. This is an encouraging finding, especially considering that we have used a simple SMC Sampler. We anticipate that further improvements in accuracy would result from using a more sophisticated $L$-Kernel, better recycling, novel proposal distributions or alternative resampling implementations \cite{Cumulative_Sum_no_sense}.
			The results for $N = 2^{31}$ are not shown. This is because for high values of $T_{SMC}$, the run-times of Metropolis-Hastings and the SMC Sampler for low values of $P$ exceed the simulation time limit on both clusters (which is set to $3$ days).
			\vspace{-10pt}
			\begin{table}[htb]
				\small
				\begin{center}
					\caption{Details of the clusters.}
					\begin{tabular}{|l|l|l|}
						\hline
						{\bf Name}       & {\bf Barkla}           & {\bf Chadwick} \\
						\hline
						{OS}	         & {CentOS Linux 7}                & {RHEL 6.10 (Santiago)}  \\
						Number of Nodes  & 16                     & 8     \\
						Cores per node   & 40                     & 16 \\
						CPU              & 2 Xeon Gold 6138       & 2 Xeon(R) E5-2660\\
						RAM              & 384 GB                 & 64 GB\\
						MPI Version      &  OpenMPI-1.10.1        & OpenMPI-1.5.3\\ 
						Max time per job & 72 hours               & 72 hours \\
						\hline
					\end{tabular}
					\label{tab:sysdetails}
				\end{center}
			\end{table}
			\vspace{-14pt}
			\begin{table}[htb]
				\caption{Metropolis-Hastings: RMSE (log scale)} 
				\label{tab: mh_rsme}
				\small
				\begin{center}
					\renewcommand{\arraystretch}{1.3}
					\begin{tabular}{| c  c | c | c | c |}
						\hline
						\multirow{2}{*}{} && 
						\multicolumn{3}{ c |}{$T_{SMC}$} \\
						\cline{3 - 5} 
						&& $10^{2}$ & $10^{3}$ & $10^{4}$ \\
						\hline 
			
						\multicolumn{1}{| c }{\multirow{3}{*}{$N$} } &
			
						\multicolumn{1}{| c |}{\multirow{1}{*}{$2^{10}$}}  
						& $-10.12$ & $-11.62$ & $-12.36$ \\ \cline{2 - 5}
			
						& \multicolumn{1}{| c |}{\multirow{1}{*}{$2^{17}$}}   
						& $-12.41$ & $-13.06$ & $-13.62$ \\ \cline{2 - 5}
			
						& \multicolumn{1}{| c |}{\multirow{1}{*}{$2^{24}$}}   
						& $-13.64$ & $-13.87$ & $-13.95$ \\ \cline{2 - 5}
			
						\hline
					\end{tabular}
				\end{center}
			\end{table}
			\vspace{-10pt}

\section{Conclusions} \label{sec:conclusions}
	In this paper, we have shown that a parallel implementation of the SMC Sampler on distributed memory architectures is an advantageous alternative to Metropolis-Hastings as it can be up to $85$ times faster over the same workload, and up to $8$ times more accurate over the same run-time for $512$ cores.
	
	To get to this position, we have made several advances. An MPI implementation of the SMC Sampler has previously been unavailable but we have proven that it can be produced by porting the key components of the Particle Filter. There exist several alternative algorithms to perform the common bottleneck, redistribute, including a state-of-the-art parallel algorithm and a textbook non-parallelisable implementation. In this paper, we have optimised the parallel algorithm and proven it can outperform the current approach for any number of cores and be up to $3$ times faster than the textbook implementation for a sufficiently high degree of parallelism. In addition, we have demonstrated the infeasibility of the non-parallelisable algorithm for large numbers of particles.
	
	The proposed algorithm for $512$ cores is $100$ times as fast as its serial configuration in the worst case scenario, which occurs when resampling (and redistribute) is needed at every step and, most importantly, when the model is unrealistically simple and hence Importance Sampling has a very fast constant time. More realistic models have highly computationally intensive proposal distributions or likelihoods. Under these realistic conditions, we have shown that the overall speed-up increases with the workload of Importance Sampling and the maximum recorded speed-up is about $254$ for $512$ cores.
	
	A key observation we can make is that the SMC Sampler version we have used is a basic reference version as the L-kernel is equal to the proposal distribution, which is Gaussian; better recycling and resampling are yet to be explored. This means that we still have left significant scope for future improvements. A combination of intelligent recycling, a more sophisticated L-Kernel, improved proposal distribution and better resampling may have major impacts on the accuracy. 
	
	Another improvement avenue is to speed up the run-time which as we have seen can indirectly improve the accuracy too. One possible way to achieve this goal is to investigate the benefits of mixing shared memory architectures and distributed memory architectures. OpenMP is the most common programming model for shared memory architectures and including OpenMP algorithms within MPI is a routine approach in the high performance computing domain. Data locality may also lead to alternative and more efficient ways of implementing redistribute. A second environment that may lead to further speed-up consists of using the additional computational power that the GPU card within each machine provides.
	
	Future work will focus on implementing all these improvements and comparing the resulting SMC Sampler with better MCMC methods than Metropolis-Hastings. These comparisons must necessarily be made both in the  single and multiple chain contexts.
\section*{Acknowledgments}
	This work was supported by the UK EPSRC Doctoral Training Award and by Schlumberger.
\appendixpage
	The following algorithms summarise the routines which are described in detail in Sections \ref{sec:smcmethods} and \ref{sec:enhancements} and compared in Section \ref{sec:evaluation} of the main paper. In these algorithms, we use the same notation of the main paper: arrays and matrices are in bold while scalars, such as some input parameters and single elements of one-dimensional arrays, are written in italic font-style.
	\section*{SMC methods and Metropolis-Hastings}
		Algorithms \ref{alg:pf}, \ref{alg:smcsampler} and \ref{alg:MH} explain the two considered SMC methods and Metropolis-Hastings. 
		\begin{algorithm}[htb]
			\caption{\label{alg:pf}SIR Particle Filter}
			\textbf{Input: } $T$, $N$, $N^*$ \\
			\textbf{Output: } $\mathbf{f}_t$
			\begin{algorithmic}[1]
				\State $\mathbf{x}_0, \mathbf{w}_0 \gets$ \texttt{Initialisation}$\left(\right)$, each particle is initially drawn from the prior distribution $q(\mathbf{x}_0) = p(\mathbf{x}_0)$ and each weight is initialised to $1/N$
				\For{$t\gets 1$; $t \leq T$; $t\gets t+1$}
					\State $\mathbf{Y}_t \gets$ \texttt{New\textunderscore Measurement}$\left(\right)$
					\State $\mathbf{x}_t, \mathbf{w}_t \gets$ \texttt{Importance\textunderscore Sampling}$()$, see (\ref{eq:pf_drawing_particles}) and  (\ref{eq:pf_drawing_weights}) in the main paper
					\State $\mathbf{\tilde{w}}_t \gets$ \texttt{Normalise}$(\mathbf{w}_t)$, see (\ref{eq:normalise}) in the main paper
					\State $N_{eff} \gets$ \texttt{ESS}$(\mathbf{\tilde{w}}_t)$, see (\ref{eq:neff}) in the main paper
					\If{$N_{eff}<N^*$}
						\State $\mathbf{x}_t, \mathbf{w}_t \gets$ \texttt{Resampling}$(\mathbf{x}_t, \mathbf{\tilde{w}}_t, N)$
					\EndIf
					\State $\mathbf{f}_t \gets$ \texttt{Mean}$(\mathbf{x}_t, \mathbf{w}_t)$, calculate the weighted mean of the particles to estimate the real state.
				\EndFor
			\end{algorithmic}
		\end{algorithm}
		\begin{algorithm}[htb]
			\caption{\label{alg:smcsampler}SMC sampler with recycling}
			\textbf{Input: } $T$, $N$, $N^*$ \\
			\textbf{Output: } $\mathbf{x}_t$, $\mathbf{\hat{f}}$
			\begin{algorithmic}[1]
				\State $\mathbf{x}_0, \mathbf{w}_0 \gets$ \texttt{Initialisation}$\left(\right)$, $\mathbf{x}_0 \sim q(\mathbf{x}_0)$ and each particle is assigned to its initial weight $w_0^i = \frac{\pi_0(\mathbf{x}_0^i)}{q_0(\mathbf{x}_0^i)}$
				\For{$t\gets 1$; $t \leq T$; $t\gets t+1$}
					\State $\tilde{c}_t \gets$ \texttt{Normalisation\textunderscore Constant}$(\mathbf{w}_t)$, see (\ref{eq: normconst}) in the main paper
					\State $\mathbf{x}_t, \mathbf{w}_t \gets$ \texttt{Importance\textunderscore Sampling}$()$, see (\ref{eq:smcs_drawing_weights}) in the main paper for $\mathbf{w}_t$; $\mathbf{x}_t \sim  q(\mathbf{x}_t|\mathbf{x}_{t-1})$
					\State $\mathbf{\tilde{w}}_t \gets$ \texttt{Normalise}$(\mathbf{w}_t)$, see (\ref{eq:normalise}) in the main paper
					\State $\mathbf{f}_t \gets$ \texttt{Estimate}$(\mathbf{x}_t, \mathbf{w}_t)$, see (\ref{eq:estimation_at_t}) in the main paper
					\State $N_{eff} \gets$ \texttt{ESS}$(\mathbf{\tilde{w}}_t)$, see (\ref{eq:neff}) in the main paper
					\If{$N_{eff}<N^*$}
						\State $\mathbf{x}_t, \mathbf{w}_t \gets$ \texttt{Resampling}$(\mathbf{x}_t, \mathbf{\tilde{w}}_t, N)$
					\EndIf
				\EndFor
				\State $ \mathbf{\hat{f}} \gets $\texttt{Recycling}$(\mathbf{f}, \mathbf{\tilde{c}}, T)$, see (\ref{eq:recycling}) in the main paper
			\end{algorithmic}
		\end{algorithm}
	
		\begin{algorithm}[htb]
			\caption{\label{alg:MH}Metropolis-Hastings}
			\textbf{Input: } $T$, $\epsilon$, $\mathbf{\Sigma}$ \\
			\textbf{Output: } $\mathbf{x}_t$
			\begin{algorithmic}[1]
				\State $\mathbf{x}_0 \sim q(\mathbf{x}_0)$
				\For{$t\gets 1$; $t \leq T$; $t\gets t+1$}
					\State $\mathbf{x}^* \sim N(\mathbf{x}^*|\mathbf{x}_{t-1}, \epsilon^2 \mathbf{\Sigma})$, a new sample is drawn from the proposal distribution 
					\State $a = min\{1,\frac{\pi(\mathbf{x}^*) q(\mathbf{x}_{t-1}|\mathbf{x}^*)}{\pi(\mathbf{x}_{t-1})q(\mathbf{x}^*|\mathbf{x}_{t-1})}\}$, calculate the acceptance probability
					\State $r \sim [0, 1]$ 
					\If{$a < r$}
						\State $\mathbf{x}_{t} = \mathbf{x}^*$, the proposed sample is accepted
					\Else
						\State $\mathbf{x}_{t} = \mathbf{x}_{t-1}$, the proposed sample is rejected and the old sample is propagated to the next iteration
					\EndIf
				\EndFor
			\end{algorithmic}
		\end{algorithm}
		
	\section*{Resampling and Redistribute}
		Algorithm \ref{alg:res} depicts the chosen resampling step for our implementation of the Particle Filter and the SMC Sampler. The three considered MPI implementations of the constituent redistribute step are described by Algorithms \ref{alg: centralised_redistribute}, \ref{alg: BSB-R} and \ref{alg: NSB-R}. These routines make use of Algorithm \ref{alg: systematic_redistribute} to redistribute within each core once the workload is balanced (or centralised as in Algorithm \ref{alg: centralised_redistribute}). Algorithm \ref{alg: Serial_Nearly_Sort} explains the single-core Nearly Sort that is used for the MPI Nearly Sort in Algorithm \ref{alg: NSB-R} which replaces the MPI Bitonic Sort in Algorithm \ref{alg: BSB-R}. 
		\vspace{-5pt}
		\begin{algorithm}[htb]
			\caption{Minimum Variance Resampling}  \label{alg:res}
			\textbf{Input: } $\mathbf{x}_t, \mathbf{w}_t, N$ \\
			\textbf{Output: } $\mathbf{x}_t, \mathbf{w}_t$
			\begin{algorithmic}[1]
				\State $\mathbf{ncopies} \gets $ \texttt{MVR}$\left(\mathbf{w}_t\right)$, apply Minimum Variance Resampling to generate $\mathbf{ncopies}$ from $\mathbf{w}_t$
				\State $\mathbf{x}_t \gets $ \texttt{Redistribute}$\left(N, \mathbf{ncopies}, \mathbf{x}_t\right) $
				\State $\mathbf{w}_t \gets $ \texttt{Reset}$\left(\mathbf{w}_t\right)$, all weights are reset to $1/N$
			\end{algorithmic}
		\end{algorithm}
		\begin{algorithm}[htb]
			\caption{Sequential Redistribute (S-R)}\label{alg: systematic_redistribute}
			\textbf{Input: } $N$, $\mathbf{ncopies}$, $\mathbf{x}$ \\
			\textbf{Output: } $\mathbf{x}_{new}$
			\begin{algorithmic}[1]
				\State $i \gets 0$
				\For{$j\gets 0$; $j < N$; $j\gets j+1$}
					\For{$k\gets 0$; $k < ncopies^j$; $k\gets k+1$}
						\State $\mathbf{x}^i_{new}\gets \mathbf{x}^i$
						\State $i\gets i+1$
					\EndFor
				\EndFor
			\end{algorithmic}
		\end{algorithm} 
		\begin{algorithm}[htb]
			\small
			\caption{Centralised Redistribute (C-R)}\label{alg: centralised_redistribute}
			\textbf{Input: } $N$, $\mathbf{ncopies}$, $\mathbf{x}$, $mpi\textunderscore rank$, $P$ \\
			\textbf{Output: } $\mathbf{x}$
			\begin{algorithmic}[1]
				\If{$mpi\textunderscore rank == 0$}
					\State Allocate memory for $N$ integers and $N$ particles
				\EndIf
				\State The master core (i.e., the core with $mpi\textunderscore rank = 0$) collects the $N/P$ elements in $\mathbf{ncopies}$ from each core using MPI\textunderscore Gather and stores them into $\mathbf{tmp\textunderscore ncopies}$
				\State The master core collects the $N/P$ particles in $\mathbf{x}$ from each core using MPI\textunderscore Gather and stores them into $\mathbf{tmp\textunderscore x}$
				\If{$mpi\textunderscore rank == 0$}
					\State $\mathbf{tmp\textunderscore x}_{new} \gets $ \texttt{Sequential\textunderscore Redistribute}$(N, \mathbf{tmp\textunderscore ncopies}, \mathbf{tmp\textunderscore x})$, see Algorithm \ref{alg: systematic_redistribute}
				\EndIf
				\State The master core scatters $\mathbf{tmp\textunderscore x}_{new}$ to the other cores using MPI\textunderscore Scatter. $\mathbf{x}$ is used as received buffer
				\State return $\mathbf{x}$
			\end{algorithmic}
		\end{algorithm} 
		\begin{algorithm}[htb]
			\small
			\caption{Bitonic Sort Based Redistribute (B-R)} \label{alg: BSB-R}
			\textbf{Input: } $ \mathbf{Node} = \left[\mathbf{ncopies}, \mathbf{x}\right]$, $N$, $P$, $n = \frac{N}{P}$\\
			\textbf{Output: } $\mathbf{x}$
			\begin{algorithmic}[1]
				\If{$P > 1$}
					\State \texttt{MPI\textunderscore Bitonic\textunderscore Sort}$\left(\mathbf{Node}, N, P\right)$
			\EndIf
			\end{algorithmic}
			\begin{algorithmic}[1]
				\Procedure{Distribute}{$\mathbf{Node}, N, P, n$}
					\If{$N == n$}, the workload is now fully balanced as each core has $n = \frac{N}{P}$ particles to copy
						\State $\mathbf{x} \gets $ \texttt{Sequential\textunderscore Redistribute}$(n, \mathbf{ncopies}, \mathbf{x})$, see Algorithm \ref{alg: systematic_redistribute}
						\State return $\mathbf{x}$ 
					\EndIf
					\State $\mathbf{csum} \gets$ \texttt{MPI\textunderscore Cumulative\textunderscore Sum}$\left(N, P, \mathbf{ncopies}\right)$, MPI\textunderscore Scan is used between the MPI nodes
					\State $pivot \gets$ \texttt{Pivot\textunderscore Calc}$\left(\mathbf{ncopies}, \mathbf{csum}\right)$, $pivot$ is the first index of $\mathbf{csum}$ such that $csum^\mathit{pivot} \geq N/2$
					\State $r \gets pivot - \left( \frac{N}{2} - 1\right)$
					\State $\left(\mathbf{Leaf_l}, \mathbf{Leaf_r}\right) \gets$ \texttt{MPI\textunderscore Rotational\textunderscore Shifts}$\left(\mathbf{Node}, r\right)$, up to $log_2P$ MPI\textunderscore Sendrecv since $r$ is expressed as a sum of power of $2$ numbers.
					\State \texttt{Distribute}$\left(\mathbf{Leaf_l}, N/2, P/2, n\right)$, the left node becomes a new father node and the size of the problem is halved
					\State \texttt{Distribute}$\left(\mathbf{Leaf_r}, N/2, P/2, n\right)$, the right node becomes a new father node and the size of the problem is halved
				\EndProcedure
			\end{algorithmic}
		\end{algorithm}
		\begin{algorithm}[htb]
			\caption{Sequential Nearly Sort (S-NS)}\label{alg: Serial_Nearly_Sort}
			\textbf{Input: } $N$, $\mathbf{ncopies}$, $\mathbf{x}$
			\\ \textbf{Output: } $\mathbf{x}_{new}, \mathbf{ncopies}_{new}$
			\begin{algorithmic}[1]
				\State $l \gets 0, r \gets N - 1$
				\For{$i\gets 0$; $i < N$; $i\gets i+1$}
					\If{$ncopies^i > 0$} 
						\State $ncopies_{new}^r \gets ncopies^i$
						\State $\mathbf{x}_{new}^r \gets \mathbf{x}^i$
						\State $r \gets r - 1$
					\Else
						\State $ncopies_{new}^l \gets ncopies^i$
						\State $\mathbf{x}_{new}^l \gets \mathbf{x}^i$
						\State $l \gets l + 1$
					\EndIf
				\EndFor
			\end{algorithmic}
		\end{algorithm}
	
		\begin{algorithm}[htb]
			\small
			\caption{Nearly Sort Based Redistribute (N-R)} \label{alg: NSB-R}
			\textbf{Input: } $ \mathbf{Node} = \left[\mathbf{ncopies}, \mathbf{x}\right]$, $N$, $P$, $n = \frac{N}{P}$\\
			\textbf{Output: } $\mathbf{x}$
			\begin{algorithmic}[1]
				\If{$P > 1$}
					\State \texttt{MPI\textunderscore Nearly\textunderscore Sort}$\left(\mathbf{Node}, N, P\right)$
				\EndIf
			\end{algorithmic}
			\begin{algorithmic}[1]
				\Procedure{Distribute}{$\mathbf{Node}, N, P, n$}
					\If{$N == n$}, the workload is now fully balanced as each core has $n = \frac{N}{P}$ particles to copy
						\State $\mathbf{x} \gets $ \texttt{Sequential\textunderscore Redistribute}$(n, \mathbf{ncopies}, \mathbf{x})$, see Algorithm \ref{alg: systematic_redistribute}
						\State return $\mathbf{x}$ 
					\EndIf
					\State $\mathbf{csum} \gets$ \texttt{MPI\textunderscore Cumulative\textunderscore Sum}$\left(N, P, \mathbf{ncopies}\right)$, MPI\textunderscore Scan is used between the MPI nodes
					\State $pivot \gets$ \texttt{Pivot\textunderscore Calc}$\left(\mathbf{ncopies}, \mathbf{csum}\right)$, $pivot$ is the first index of $\mathbf{csum}$ such that $csum^\mathit{pivot} \geq N/2$
					\State $r \gets pivot - \left( \frac{N}{2} - 1\right)$
					\State $\left(\mathbf{Leaf_l}, \mathbf{Leaf_r}\right) \gets$ \texttt{MPI\textunderscore Rotational\textunderscore Shifts}$\left(\mathbf{Node}, r\right)$, up to $log_2P$ MPI\textunderscore Sendrecv since $r$ is expressed as a sum of power of $2$ numbers.
					\State \texttt{Distribute}$\left(\mathbf{Leaf_l}, N/2, P/2, n\right)$, the left node becomes a new father node and the size of the problem is halved
					\State \texttt{Distribute}$\left(\mathbf{Leaf_r}, N/2, P/2, n\right)$, the right node becomes a new father node and the size of the problem is halved
				\EndProcedure
			\end{algorithmic}
		\end{algorithm}	

\bibliographystyle{unsrt}
\bibliography{bibs}

\end{document}